\documentclass[tradiabstract]{aa} % for the abstract without structuration
\usepackage{graphicx}
\usepackage{float}
\usepackage{txfonts}
\usepackage[breaklinks, colorlinks, citecolor=blue]{hyperref}
\usepackage{natbib}
\bibpunct{(}{)}{;}{a}{}{,} % to follow the A&A style
\newcommand{\vn}{\mbox{\bf {n}}}

\newcommand{\vs}{{\bf {s}}}

\newcommand{\vl}{{\bf {l}}}

\newcommand{\vv}{\mbox{\bf {v}}}

\usepackage{multirow}

\begin{document}
   \title{Distortions of the Cosmic Microwave Background through cooling lines during the epoch of Reionization}
   % \subtitle{Sensitivity of Thomson scattering on the local CMB monopole}
   % \subtitle{I. Overviewing the $\kappa$-mechanism}
   \author{
		Carlos Hern\'andez-Monteagudo\inst{1},
    		Umberto Maio\inst{2},
		Benedetta Ciardi\inst{3},
       and
       	Rashid A. Sunyaev\inst{3,4}		
		 \fnmsep
	}
   \institute{
		Centro de Estudios de F\'\i sica del Cosmos de Arag\'on (CEFCA),
   		Plaza San Juan 1, planta 2, E-44001 Teruel, Spain\\
		\email{chm@cefca.es}
       		\and
		Leibniz Institute for Astrophysics Potsdam (AIP),
		An der Sternwarte 16, D-14482 Potsdam, Germany\\
		\email{umaio@aip.de}
		\and
        	Max Planck Institut f\"ur Astrophysik,
        	Karl Schwarzschild Str. 1, D-85741 Garching bei M\"unchen, Germany\\
        	\email{ciardi,sunyaev@mpa-garching.mpg.de}
        		\and
        	Space Research Institute, 117810 Moscow, Russia
	}

   \date{Received ; accepted }

% \abstract{}{}{}{}{} 
% 5 {} token are mandatory
 
  \abstract{
 By using N-body hydrodynamical cosmological simulations in which the chemistry of major metals and molecules is consistently solved for, we study the interaction of metallic fine-structure lines with the cosmic microwave background radiation (CMB). Our analysis shows that the collisional induced emissions in the OI 145\,$\mu$m and CII 158\,$\mu$m lines during the epoch of reionization ($z>5$) introduce a distortion of the CMB black-body spectrum at low frequencies ($\nu < 300$\,GHz) with amplitudes up to  $\Delta I_{\nu}/B_{\nu}(T_{\rm CMB})\sim 10^{-8}$--$10^{-7}$, i.e., at the $\sim 0.1$ percent level of FIRAS upper limits. Shorter wavelength fine-structure transitions (like OI 63\,$\mu$m, FeII 26\,$\mu$m, SiII 35\,$\mu$m, FeII 35\,$\mu$m, and FeII 51\,$\mu$m) typically sample the reionization epoch at higher observing frequencies ($\nu\in[400,1000]$\,GHz). This corresponds to the Wien tail of the CMB black-body spectrum and thus the distortion level induced by those lines may be as high as $\Delta I_{\nu}/B_{\nu}(T_{\rm CMB})\sim 10^{-4}$. Consequently, the brightness temperature anisotropy produced by these lines should be more relevant at higher frequencies: while practically negligible at $\nu=145\,$GHz, signatures from CII 158\,$\mu$m and OI 145\,$\mu$m should amount to 1\,\%--5\,\% of the anisotropy power measured at $l \sim 5000$ and $\nu=220\,$GHz by the ACT and SPT collaborations (after taking $\Delta \nu_{\rm obs}/\nu_{\rm obs}\simeq 0.005$ for the line observations). More importantly, our simulations indicate that anisotropy maps from different lines (e.g., OI 145\,$\mu$m and CII 158\,$\mu$m) at the same redshift show a very high degree ($>0.8$) of spatial correlation, allowing for the use of observations at different frequencies (under different systematic and noise contributions) to unveil the same snapshot of the reionization epoch. Finally, our simulations also demonstrate that line-emission anisotropies extracted in narrow frequency/redshift shells are practically uncorrelated in frequency space, thus enabling standard methods for removal of foregrounds that vary smoothly in frequency, just as in HI 21\,cm studies.
}
%   {investigate.}
%   {It is shown that }
%   {The stability equations }
 %  {Vibrational instability }
 %__ traditional abstract
 
   \authorrunning{Hern\'andez-Monteagudo et al.}
   \titlerunning{CMB distortions induced by cooling lines during reionization}
   \maketitle
%
%________________________________________________________________

\section{Introduction}

Exploring the epoch of cosmological reionization constitutes one of the standing challenges in observational cosmology.
Indeed, there is a number of questions motivating nowadays tremendous efforts to access those early stages in the history of the universe: when did the first stars and proto-galaxies appear in the universe? Which sources were actually responsible for reionizing the intergalactic medium (IGM)? How was the assembly of the large scale structure modified by feedback effects associated to radiation? How was the IGM polluted with heavy elements? In order to address these and other questions there exist currently a few observational  windows attempting to access the epoch of cosmological reionization.
Probably the most promising of them is the 21\,cm detection of the hyperfine transition of neutral hydrogen. Those 21\,cm photons, when emitted at the epoch of reionization, get redshifted deep into the radio domain, and provide information on the different mechanisms causing the spin temperature to depart from the temperature of the ambient radiation field, namely the cosmic microwave background radiation (CMB), \citep[see, e.g., ][for a detailed review on the subject]{pritchard12}.
Different physical phenomena, such as the adiabatic cooling of baryons, the collisions between hydrogen atoms with electrons and other atoms, or the coupling of hydrogen with Lyman-$\alpha$ photons \citep{wout, field58} make the spin temperature of the 21\,cm transition detach from the CMB one, causing such transition to be seen either in emission or in absorption with respect to the CMB background.
Experiments like 
LOFAR\footnote{\tt http://www.lofar.org/} \citep[][]{lofar}, 
MWA\footnote{\tt http://www.mwatelescope.org/} \citep[][]{mwa} 
and SKA\footnote{\tt https://www.skatelescope.org/} \citep[][]{ska} 
are attempting to map the onset of reionization through this hyperfine transition.
While their sensitivity is sufficient for the expected signal amplitude, these experiments still need to keep radio foregrounds, of larger amplitude by several orders of magnitude, under control.

Another window to the epoch of reionization is provided by the rest-frame UV and optical radiation from the reionization sources, which becomes redshifted in the near-mid infrared range for local observers. Medium and narrow band surveys of the type of MUSE\footnote{\tt http://muse-vlt.eu/science/}, the SUBARU Deep Field \citep[][]{subaruLAEs} and SHARDS\footnote{\tt https://guaix.fis.ucm.es/\~\,pgperez/SHARDS/} have already detected a significant amount of high-redshift ($z\in [5,7]$) Lyman-$\alpha$ blobs, and currently further studies are conducted in order to estimate luminosity functions and total ionizing flux from those sources. The imminent launch of the {\it James Webb Space Telescope}\footnote{\tt http://www.jwst.nasa.gov/} will open a new era in this type of observations.

In this work we focus on a different approach to access the epoch of reionization. The CMB photons cross the entire observable universe before reaching us and, in particular, they witness the birth of the first stars and galaxies and all the subsequent structure formation and evolution. The CMB should hence carry some signatures from the epoch of reionization. As reviewed in e.g. \cite{jens_review_16}, the leading signature on the CMB spectrum from the epoch of reionization should correspond to the inverse Compton scattering \citep[via the thermal Sunyaev-Zeldovich effect,][hereafter tSZ]{tSZ} off hot electrons. This tSZ should leave a $y$-type distortion in the CMB black-body spectrum with an amplitude of $y\simeq 2\times 10^{-6}$ \citep[see, for more details][]{Hill2015}. Other processes potentially distorting the CMB black-body spectrum during the epoch of reionization are the collisional emission on fine-structure lines of high-$z$ metals \citep{suginohara2spergel99} or the admixture of heavy elements with the primeval gas \citep{vashalovichetal81}. 

However, not all physical processes during reionization introduce a distortion on the CMB black-body. As first noted by \citet{basu04}, resonant scattering should be the main interaction channel between CMB photons and metals and molecules produced by stars, and/or in the surroundings thereof, since collisions should only be efficient in very large overdense regions ($\delta \gtrsim 10^{5}$). Provided that resonant scattering does not change the number of CMB photons (thus preserving the CMB spectrum), \citet{basu04} computed the impact of heavy-element fine-structure resonant scattering on the CMB intensity anisotropies, after assuming a uniform distribution for cosmic metals. They found that the leading effect was a blurring of the intrinsic CMB anisotropies. The resulting small-scale variations in the CMB angular power spectrum could be approximated by $\Delta C_l \simeq -2\tau C_l^{\rm CMB,\, intr}$, with $C_l^{\rm CMB,\,intr}$ intrinsic CMB angular power spectrum and $\tau$ optical depth associated to the given transition\footnote{
We remind that the optical depth scales linearly with the number density of the resonant species.
}.
This idea was further explored by \citet{chm_metals06, chm_pol_07} and then extended to study {\em both} spectral distortion {\em and} angular anisotropies associated to the Field-Wouthuysen effect arising in OI atoms during the epoch of reionization \citep{chm_OI_I, chm_OI_II}. Although this effect does not exclusively apply to OI atoms in extremely overdense regions, the amplitude of the distortions foreseen for this mechanism was relatively modest ($\sim 10^{-9}$--$10^{-8}$ ).

In the present work we explore distortions and angular anisotropies induced by several fine-structure lines associated to metal spreading during the early stellar evolutionary stages in the epoch of reionization. This constitutes the numerical implementation of the early estimates of 
\citet{suginohara2spergel99}.
We also improve the work of \citet{basu04} in the sense that, instead of assuming a uniform distribution of metals in the Universe, we model their clustering by looking at the outputs of state-of-the-art N-body hydrodynamical chemistry simulations, consistently including metal spreading and corresponding radiative losses from resonant and fine-structure transitions of individual heavy elements.
This allows us to track precisely the impact of collisionally-induced emission on a set of fine-structure transitions (which do distort the CMB black-body spectrum) on top of the resonant scattering addressed in \citet{basu04}.

We find that the distortion and angular anisotropies are essentially induced by the resonant scattering and collisional emission on fine-structure transitions. The former {\em blurs} the intrinsic CMB anisotropies from large to intermediate angular scales \citep[the CMB angular power spectrum changes as $\Delta C_l^{\rm blur} \simeq -2\tau_{X,\,\nu}^{ij}\, C_l^{\rm CMB,\,intr}$, with $\tau_{X,\,\nu}^{ij}$ optical depth associated to a given transition $ij$ of the $X$ species, see ][]{basu04}. The impact of the peculiar velocity of the scatterers (the so-called Doppler term) adds practically negligible power on all angular scales. On the other hand, the collisional emission dominates on small angular scales. The relative amplitude of the collisional emission with respect to the CMB increases with the spectral resolution of the observations \citep[in agreement with][]{righi08}, and it also increases with higher frequencies sampling the later stages of the epoch of reionization. At and above $\nu_{\rm obs}=220$\,GHz we find that, for a spectral resolution of $\Delta \nu_{\rm obs}/\nu_{\rm obs}=5\times 10^{-3}$, the collisional emission term surpasses the impact of blurring (due to resonant scattering) on practically all angular scales, contributing to $\gtrsim 10$\,\% of the anisotropy power at $l>2000$. 

This paper is structured as follows: in Sect.~\ref{sec:sims} we describe the hydrodynamical numerical simulations under use; in Sect.~\ref{sec:interac} we outline the different terms of the radiative transfer equation for CMB photons and its integration. In Sect.~\ref{sec:results} we describe the results in terms of predictions for the distortion of the CMB spectrum and the angular clustering of the line-induced anisotropies. Finally, in Sect.~\ref{sec:discon} we discuss our results and their implications in the context of upcoming CMB-distortion and HI 21\,cm experiments and conclude.

\section{The simulations}

\label{sec:sims}

Our work is based on the output of hydrodynamical cosmological numerical simulations. From the initial stages laid down by seminal works \citep{holmberg41,vonhoerner60,vonhoerner63,barnesandhut86,hockneyandeastwood88,cenostriker90,cenostriker93,gnedin96I,gnedin96II,gnedinbertschinger96}, numerical techniques have evolved and been refined over the years until becoming a standard tool in cosmology \citep{gadget}. 
The numerical calculations performed in this work extend our previous studies of the intergalactic medium during the epoch of reionization.
We employ the outputs of the hydrodynamical simulations presented in \citet{Maio2010} and  \citet{ Maio2011}.
Besides gravity and hydrodynamics calculations, the implementation contains a self-consistent treatment of molecules and metals, too, as discussed in \citet{maio07}.
We include detailed non-equilibrium atomic and molecular chemistry evolution for electrons,
H, 
H$^+$, 
H$^-$, 
He, 
He$^+$, 
He$^{++}$, 
D, 
D$^+$, 
H$_2$, 
H$_2^+$, 
HD, 
HeH$^+$.
The variations of the number density of individual species are computed according to a backward differential scheme. The adopted timestep for the solution of the differential equations is 1/10th the hydrodynamic timestep. This choice assures a robust convergence of our calculations, as previously checked in the literature by, e.g., \citet{Anninos1997, Yoshida2003, Maio2013, Maio2016}.

Star formation and stellar evolution is followed to track stellar lifetimes and metal-dependent yields from different types of stars, including C, N, O, Ne, Mg, Si, S, Ca, Fe \citep{Tornatore2010}. The  corresponding resonant cooling and fine-structure transition cooling below $10^4\,$K (relevant for this work) is also monitored \citep[see][for details]{maio07}. In this way, cooling from sub-mm fine-structure transitions is treated fully self-consistently with the thermal gas evolution.
To estimate the fine-structure emissions consistently with the simulation hydrodynamics, we compute at runtime the atomic level populations by relying on particle conservation and the detailed balance principle in the low-density limit (level populations treated in this way naturally converge also at high densities).
Cooling rate emissions are then derived through the level populations, the radiative transition probabilities and the corresponding energy separations. % \citep{maio07}.

As sources of metal pollution we consider both population III (popIII) and population II-I (popII-I) stellar generations with corresponding stellar yields for 
massive SNe \citep{WW1995, WH2002},
AGB stars \citep{vdHoek1997}
and SNIa events \citep{Thielemann2003}.
The assumed stellar initial mass function covers the range $[0.1, 100]\,$M$_{\odot}$ for popII-I stars and $[100, 500]\,$M$_{\odot}$ for popIII stars.
The energy of massive popII-I SNe is assumed to be $\sim 10^{51}\,$erg, while the energy of massive popIII pair-instability SNe is mass-dependent and reaches $\sim 10^{53}\,$erg \cite[see more details and discussions in e.g.][]{Maio2010}.
The transition from early popIII to subsequent popII-I generations is determined by a critical metallicity $ Z_{\rm crit}= 10^{-4}\,$Z$_{\odot}$ \citep{Bromm2003, Schneider2006}.
For gas with $ Z < Z_{crit}$ ($>Z_{\rm crit}$) popIII (popII-I) star formation is adopted.
Popular stellar lifetimes are derived by \citet{MatteucciGreggio1986} and \citet{RenziniBuzzoni1986}.
Metal diffusion in the cosmic medium is mimicked by smoothing individual metallicities over the neighbouring particles in the SPH kernel.
A star formation density threshold of 1 particle per cubic cm is used, because this is the minimum value required to obtain converging and meaningful results at early epochs, both in terms of star formation rate \citep[for a deeper analysis see e.g. ][]{Maio2009} and of chemical patterns \citep{MT2015}.
We refer to \citet{Maio2013} for results of more extended tests on the effects of proper inclusion of cooling, metal spreading and a full parameter exploration.

Throughout this paper we adopt the following cosmological parameters:
$H_0 = 70$\,km\,s$^{-1}$\,Mpc$^{-1}$, 
$\Omega_{\rm m} = 0.3$, 
$\Omega_{\rm b} = 0.04$, 
$\Omega_{\Lambda} = 0.7$, and $n_S=1, \sigma_8=0.9$ for the scalar spectral index and the current amplitude of linear matter fluctuations in spheres of 8\,$h^{-1}$\,Mpc radius, respectively.

\section{The interaction of the CMB with metals during reionization}
\label{sec:interac}

We consider the metal species C, O, Si and Fe and their interaction with the CMB via a set of fine-structure transitions in both neutral and ionized states of the atoms. The two channels for this interaction considered here are the resonant scattering and the collisional emission on those transitions.

Metal fine-structure transitions constitute a source of optical depth for CMB photons, which can be scattered resonantly and hence have their direction of propagation changed. The effect of this process is the {\it blurring} of the original small-scale ($l>20-50$) CMB anisotropies and the generation of new anisotropies {\it if} the scatterers have a peculiar motion with respect to the CMB photon field. This mechanism of generation of new anisotropies will be referred to as the {\it Doppler} term.
%%%%
Resonant scattering does not distort the CMB black-body spectrum, since no new photons are produced.

At the same time, if metals are placed in a high-density environment, then collisions with neutral hydrogen atoms may excite them into upper levels of the fine-structure transitions considered. The subsequent de-excitation to the lower levels would result in the emission of a photon at the transition frequency. This process of collisional emission produces {\it new} photons that do distort the CMB black-body spectrum, unlike resonant scattering.

Another process which introduces a distortion is the UV pumping of the resonant transitions. This physical process, also known as the Wouthuysen-Field effect \citep{wout, field58}, has been extensively studied when characterizing the H 21 cm fluctuations during the epoch of reionization, and also in the context of metal emission during that same epoch \citep{chm_OI_I,  chm_OI_II, kawasaki12}. The presence of a UV radiation field causes excitation in the electronic metallic levels and this process itself modifies the spin temperature of the fine-structure transition(s). Nevertheless, to properly account for this effect, one requires a precise knowledge of the spectrum of the local UV field, a problem that goes beyond the scope of this work. We foresee, however, that the amplitude of this distortion is significantly smaller than the collision-induced distortion addressed in this work. 

In what follows, we describe how we address resonant scattering and collisional emission in our simulations.
We remind that the interaction of CMB photons with atoms and molecules moving along a direction $\bf n$ can be described by the radiative transfer equation modelling the evolution of the specific intensity at frequency $\nu$, $I_\nu$, in terms of proper distance $\bf s$: 
\begin{equation}
	\frac{dI_{\nu}(\vn,\vs)}{ds} = -\dot{\tau}^{ij}_{X,\,\nu} (\vs)\biggl( I_{\nu}(\vn,\vs)  -I^0_{\nu} (\vs)\biggr) + j_{\nu}(\vn,\vs).
\label{eq:radT1}
\end{equation}
In this equation, effects of the Hubble expansion have been neglected as our simulated boxes are typically much smaller than the Hubble radius at the redshifts of interest (this approximation is commonly adopted in these cases). For the sake of simplicity, the opacity $\dot{\tau}^{ij}_{X,\,\nu}$ refers only to one fine-structure transition linking levels $i\rightarrow j$ of species $X$.
The first term in the right-hand side of the equation refers to resonant scattering and is discussed in the next  Sect.~\ref{sec:interac.one}.
The second term, $j_{\nu}(\vn,\vs)$, refers to the emissivity that in our case is assumed to be caused by collisional emissions and will be discussed in Sect.~\ref{sec:interac.two}.
The optical depth, $\tau_{X,\,\nu}^{ij}$, relates to the opacity, $\dot{\tau}_{X,\,\nu}^{ij}$, via the following integral over proper distance:
\begin{equation}
\tau_{X,\,\nu}^{ij} = \int ds\, \dot{\tau}_{X,\,\nu}^{ij}.
\label{eq:opt_taudot}
\end{equation}

The relative amplitude of the intrinsic CMB intensity anisotropies is very small ($\sim 10^{-5}$) and this makes the blurring term consequently small, too.
However, as we show next, scatterers moving with respect to the CMB rest frame can make the resonant scattering term significantly larger.
Let us assume that the scatterer moves at a velocity $\vv$ with respect to (wrt) the observer (in units of the speed of light).
A CMB photon seen at frequency $\nu$ by an observer at rest wrt the CMB will have a frequency 
$\nu'\simeq \nu(1+\vv\cdot\vn)$ by the moving scatterer.
Both observers will, however, see the same CMB monopole.
By noting that $I_{\nu}/\nu^3$ is a relativistic invariant, one can easily find that the term $ I_{\nu}(\vn,\vs) -I^0_{\nu}(\vs)$ in Eq.~\ref{eq:radT1} can be rewritten, to leading order in $\vv\cdot\vn$, as
$I^0_{\nu}(\alpha-3)\vv\cdot\vn + \delta I^{\rm intr}_{\nu}(\vn,\vs)$, 
with $\alpha = \partial \log(I_{\nu}(\vn,\vs)) / \partial \log\nu$ expressing the relative variation of the specific intensity versus the relative change in frequency, and $\delta I^{\rm intr}_{\nu}(\vn,\vs)$ denoting the intrinsic CMB anisotropies.
Therefore, to first order in $\vv\cdot\vn$, we can write:
\[
\frac{dI_{\nu} (\vn,\vs)}{ds}   \simeq  -\dot{\tau}_{X,\,\nu}^{ij} (\vs)\,\, \bigg( \delta I^{\rm intr}_{\nu}(\vn,\vs) + I^0_{\nu} (\alpha-3)\vv\cdot\vn  \biggr)
\]
\begin{equation}
\phantom{xxxxxxxxxxxxxxxx}
+ j_{\nu} (\vn,\vs) [1+(\alpha-3)\vv\cdot\vn].
\label{eq:RT2}
\end{equation}
The last term on the right-hand side is the emissivity corrected by the Doppler effect due to the scatterer's velocity.
In what follows we shall neglect this Doppler corrections in the term proportional to the emissivity (i.e. the scatterer's velocity is assumed to be much smaller than the speed of light), so the equation finally reads
\begin{equation}
\frac{dI_{\nu} (\vn,\vs)}{ds} \simeq -\dot{\tau}_{X,\,\nu}^{ij} (\vs)\,\,  \biggl( \delta I^{\rm intr}_{\nu}(\vn,\vs) + I^0_{\nu}\vv\cdot\vn\,(3-\alpha)\biggr) + j_{\nu} (\vn,\vs).
\label{eq:RT3}
\end{equation}
Hereafter, we will refer to the first term in parentheses of the right-hand side of this equation as the {\it blurring term}, since it refers to the blurring of the intrinsic CMB anisotropies.
The second term, proportional to $\vv\cdot\vn$, will be denoted as the {\it Doppler term}.
Finally, the third term, proportional to the emissivity, is the {\it emission term} (see next section).
We stress that the former two terms are purely due to resonant scattering and conserve the number of photons. The third term, instead, is responsible for the distortion of the CMB black-body spectrum, since it involves the production of new photons.

After integrating Eq.~\ref{eq:RT3} above, the resulting specific intensity $I_{\nu}$ at the resonant redshift is projected onto a local observer ($z=0$) by imposing that $I_{\nu}/\nu^3$ is a relativistic invariant.  We next describe the contribution from the blurring and Doppler terms, and the collisionally-induced emission term in more detail. 
%{\bf BC: something is missing here!}

\subsection{Resonant scattering: blurring and Doppler terms}
\label{sec:interac.one}

The opacity can be written in terms of the resonant wavelength of the transition, $\lambda$, the Einstein $A_{ji}$ coefficient, the number density of the resonant species ($n_X$) and the fraction of atoms/molecules populating the considered energy level $f_l$ \citep[e.g.,][]{basu04}:
\begin{equation}
\dot{\tau}^{ij}_{X,\,\nu} = \frac{\lambda^2}{8\pi} \psi(\nu) A_{ji} \frac{g_j}{g_i} n_X(\vs) f_l. 
\label{eq:opac1}
\end{equation}
The degeneracy factors $g_i$ and $g_j$ depend on the particular transition, and the absorption frequency profile of the transition, $\psi(\nu)$, obeys to the convolution of the intrinsic absorption profile and the observational frequency profile.
In most practical situations, the latter is always much broader than the former, so we shall model $\psi(\nu)$ with a square function centred upon the resonant frequency, $\nu_X^{ij}$:
$\psi(\nu) = 1/(\Delta \nu_{\rm obs} (1+z_r))$ 
if $ \left| \nu - \nu_X^{ij} \right| < \Delta \nu_{\rm obs}(1+z_r)/2$ 
and $\psi(\nu)=0$ otherwise, 
with $\nu_x^{ij} = \nu_{\rm obs,\,c} (1+z_r)$, 
$\nu_{\rm obs,\,c}$ the central observing frequency and 
$z_r$ the resonant redshift 
connecting $\nu_{\rm obs,\,c}$ and the resonant frequency $\nu_{X}^{ij}$.
The quantity $\Delta \nu_{\rm obs} (1+z_r)$ represents the effective frequency width at the resonant epoch, and is nothing but the observational (instrumental) frequency width ($\Delta \nu_{\rm obs}$) projected to the scattering redshift.
For the emissivity $j_{\nu}$ we will assume a similar frequency profile, given by the same instrumental frequency response.

The instrumental frequency width $\Delta \nu_{\rm obs}$ conditions the width of the redshift shell in which a given experiment will be sampling the species interacting with the CMB photons. The width of this shell is given, in physical units, by
\begin{equation}
\Delta s = c H^{-1}(z) \frac{\Delta z}{1+z} =  c H^{-1}(z) \frac{\Delta \nu_{\rm obs}}{\nu_{\rm obs}},
\label{eq:Ds}
\end{equation}
provided that $d\log{\nu_{\rm obs}} = -d\log{(1+z)}$.
Formally, one should integrate Eq.~\ref{eq:RT3} above within this width only.
In order to use the full length of the box, and unless otherwise specified, $\Delta \nu_{\rm obs}$ is chosen such that $\Delta s$ equals the box size in physical units.

For the CII 158\,$\mu$m transition, this results in an effective observational frequency width ranging from $\Delta \nu_{\rm obs}\sim 0.5$\,GHz up to $\sim 1.4$\,GHz at $z=49$ and $5$, respectively,
and for the OI 63\,$\mu$m line $\Delta \nu_{\rm obs}\in [1.2,\, 3.5]\,$GHz.
For all lines, we have that $\Delta \nu_{\rm obs}/ \nu_{\rm obs}\in [5\times 10^{-3},1.3\times 10^{-2,}]$, since the relative width only depends on the resonant redshift.

\citet[][]{basu04} first predicted the signatures of resonant scattering on metals on CMB anisotropies through its linear contribution, corresponding to the blurring term, by assuming a uniform distribution of heavy elements in the Universe.
Since our simulations are able to track the level populations for several fine-structure transitions and in different cosmological environments, we can provide more realistic estimates of the average optical depth associated to them, and, consequently, we can improve the predictions on the blurring term, too.
It can be easily seen from Eq.~\ref{eq:RT3} that, on large scales, the contribution of the blurring term reduces approximately to $\Delta I^{\rm blur}_{\nu}(\vn,\vs) \simeq -\bar{\tau}_{X,\,\nu}^{ij} \delta I^{\rm intr}_{\nu}(\vn,\vs)$, where $\bar{\tau}_{X,\,\nu}^{ij}$ is the spatially averaged optical depth of the transition under study\footnote{
In what follows we shall adopt {\em average} values of the optical depth unless otherwise made explicit.
}.
In Fourier space, this last expression may be re-written as 
$\Delta I_{\nu}(\vl) \simeq -\bar{\tau}_{X,\,\nu}^{ij} \delta I^{\rm intr}_{\nu}(\vl)$, 
with $\vl$ the Fourier wave-vector such that the intrinsic CMB angular power spectrum is given by
$C_l^{\rm CMB, intr} = \langle |\delta I^{\rm intr}_{\nu}(\vl)|^2 \rangle_{\hat{\vl}}$, 
where $\langle ...\rangle_{\hat{\vl}}$ denotes the angular average over different orientations of the wave-vector $\vl$.
From this it is straightforward to find that the modification of the power spectrum due to this blurring effect obeys to $\Delta C^{\rm blur}_l = -2\tau_{X,\,\nu}^{ij} C_l^{\rm CMB, intr}$ \citep[][]{basu04}.

While on very large scales it is possible to predict the scale and redshift dependence of the gas peculiar velocity, our hydrodynamical simulations are restricted to midly to highly non-linear scales. We therefore rely on the output of these numerical simulations to track the spatial modulation of the projected peculiar momentum giving rise to the Doppler term.

\subsection{Collisional emission}
\label{sec:interac.two}

The fraction of species $X$ populating different atomic levels ($f_l$) depends on the collisional rate, which, therefore, influences the emissivity ($j_{\nu}$) of a given transition.
These quantities are {\it a priori} functions of the density of metals, electrons, protons, the UV radiation field and gas temperature.
The outputs of the numerical simulations introduced earlier have been used to tabulate the atomic relative populations in the different levels 
%  $f_l$ 
in terms of gas temperature, density and species abundance.
This allows us to obtain, for each transition, the corresponding value of $j_{\nu}(\vn,\vs)$ emitted into the surrounding medium (i.e. the amount of photons which distorts the CMB spectrum).

To estimate the fine-structure emissions consistently with the hydrodynamics of the simulated cosmic gas, we implement the energy rate density arising from collisions due to enriched material (within $10-10^9\,$K) in the energy equation of the simulation code.
At each timestep and for each gas particle, we get gas density, temperature and chemical composition as resulting also from the previous stellar evolution and pollution history.

Atomic level populations are computed by assuming particle conservation and the detailed balance principle in the low-density limit (level populations treated in this way naturally converge at high densities, too) to account for excitations and de-excitations of the considered metal species.
When a single transition of a species (like CII) is followed it is possible to solve for the two level occupation fractions analytically. For more complex species (such as OI or FeII), we need to resolve a linear system, as detailed in \citet{maio07}.
Usually, hydrogen collisions dominate the interactions of metals with cosmic gas. However, for sake of completeness, we also consider the contribution due to possible free electrons, that could be important in the presence of recent star formation or feedback effects.
The radiative losses from each transition are obtained by multiplying the upper level populations with the corresponding radiative transition probabilities and energy separations \cite[listed in Appendix B of][]{maio07}.
In this way, the resulting level populations and the corresponding fine-structure emissions are consistent with the features of the local environment and its chemical composition.

The average collisional emission induced by the transitions associated to the metals introduces an average distortion of the CMB black-body spectrum.
At any redshift $z$, this is quantified through the CMB spectral-distortion parameter, $y$, which is defined as the ratio of the collisionally-induced ``extra'' emission $\Delta I_{\nu}^{\rm coll}$ and the CMB black-body specific intensity $B_{\nu}(T_{\rm CMB}[z])$ evaluated at the corresponding resonant frequency:
\begin{equation}
	y =  \frac{ \Delta I^{\rm coll}_{\nu} }  { B_{\nu}(T_{\rm CMB}[z_r]) }.
\end{equation}
The numerator in this equation is simply the integral of Eq.~\ref{eq:RT3} that is triggered by the emissivity $j_{\nu}$,
\begin{equation}
\Delta I^{\rm coll}_{\nu} (\vn) = \int ds\, j_{\nu} (s,\vn).
\label{eq:I_col}
\end{equation}
We remark that collisional emission, unlike resonant scattering, {\em adds} photons to the CMB black-body spectrum and, thus, is the only contribution that survives after averaging over sufficiently large volumes (in which the blurring and the Doppler terms cancel at first order).
For this reason, in the following, we shall use $\Delta I^{\rm coll}_{\nu}$ and $\Delta I_{\nu}$ indistinctly when referring to spatially-averaged specific intensity.

\section{Results}
\label{sec:results}

In this section we discuss the impact that high-redshift metal lines have on the CMB.
We consider the following fine-structure lines:
CII 157.7 $\mu$m, 
SiII 34.8 $\mu$m, 
OI 63.18 $\mu$m,
OI 145.5 $\mu$m, 
FeII 25.99 $\mu$m, 
FeII 35.35 $\mu$m and
FeII 51.28 $\mu$m.
These wavelengths correspond to emitted frequencies of about 
1901, 
8615, 
4745, 
2060, 
11535,
8481 and
5846 GHz, 
respectively.
For the sake of clarity, in the following these transitions will be referred to as:
CII 158 $\mu$m,
SiII 35 $\mu$m, 
OI 63 $\mu$m, 
OI 145 $\mu$m, 
FeII 26 $\mu$m, 
FeII 35 $\mu$m and 
FeII 51 $\mu$m.

\subsection{Evolution throughout reionization}

%-----------------------
\begin{figure*}
\centering
\includegraphics[width=18.cm]{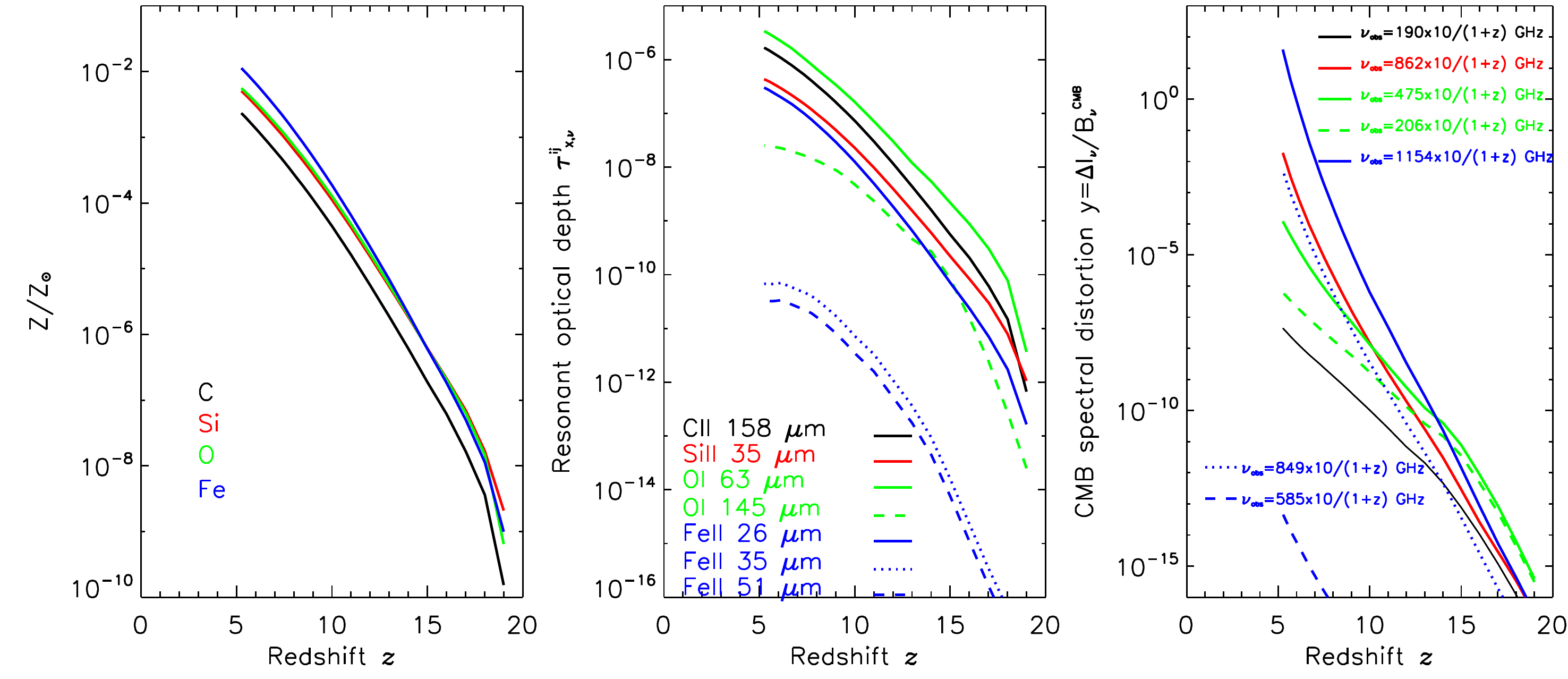}
\caption[fig:fig1-3panels]{
{\it Left panel:} Evolution of metallicity versus redshift for C (black), Si (red), O (green) and Fe (blue) abundances in solar units.
{\it Middle panel:} Optical depth associated to each of the resonant transitions considered in this work, i.e.: 
CII 158 micron (black solid line),
SiII 35 micron (red solid line),
OI 63 micron (green solid line),
OI 145 micron (green dashed line),
FeII 26 micron (blue solid line),
FeII 35 micron (blue dotted line),
FeII 51 micron (blue dashed line).
{\it Right panel:} Spectral distortions introduced by collisionally excited emission on the different transitions considered in this paper, given by the ratio of the average collisionally induced emission over the CMB intensity: $\Delta I_{\nu}/B_{\nu}(T_{\rm CMB}[z])$. Each redshift corresponds to an observing frequency, and the relation between these two is indicated by the legends. The color and linestyle coding is identical to the one in the middle panel. The observed frequencies quoted in the right panel, $\nu_{obs}$, are obtained by dividing the emitted frequencies by $(1+z)$. To better display their magnitude, they have been written by highlighting a factor of 10 for $(1+z)$, being $z$ in the range $\rm \sim [5, 20]$.

}
\label{fig:fig1-3panels}
\end{figure*}
%-----------------------

As structure formation and evolution proceed, the abundance of metals as they are produced by stars increases. As shown in the left panel of Fig.~\ref{fig:fig1-3panels}, species like carbon, silicon, oxygen or iron increase from a level close to $\sim 10^{-8}$\,Z$_{\odot}$ at $z\simeq 20$ up to $\sim 10^{-2}$\,Z$_{\odot}$ at $z\simeq 5$, 
where solar metallicities have been computed following the solar reference number abundances of \citet[][]{asplund09}.  O, C, Si and alpha elements in general come from early SNe, exploding at $z\sim 20$. After a few $10^8\,yr$ from the initial star formation phases (i.e. already at redshift $z\sim 10-15$), AGB stars contribute (mostly) with additional C, N and O. Fe can be produced by short-lived massive SNe as well as by low-mass SNIa, effective after about 1~Gyr from the onset of star formation, i.e. at $z\lesssim 6$. With the occurrence and evolution of metals in the Universe, the resonant scattering and the emissivity terms result consequentially affected.
Indeed, the optical depth of the transitions associated to these species increases accordingly to metallicities, as given in the middle panel of Fig.~\ref{fig:fig1-3panels}.
The legend in the panel relates each fine-structure transition to a line color and style. These apply to the right panel as well, where the $y$ parameter for the average collisional CMB distortion induced by each line emission is displayed. Also the distortion parameter $y$ is shown against the cosmological redshift, $z$. Additionally, the legends quote the scaling with $z$ of the observing frequency for each transition. For instance, for the CII 158\,$\mu$m transition, an observing frequency of 190\,GHz would be required to probe the emissivity signal at $z=9$, while, more generally, $\nu_{\rm obs}=190\times 10/(1+z)\,$GHz is required at any arbitrary redshift $z$. We find that the average collisional emission induced by the transitions associated to these metals grows in amplitude with decreasing redshift and causes an increasingly larger average distortion of the CMB black-body spectrum. 
This is essentially a consequence of both the larger pollution level and the smaller CMB intensity at lower $z$ (see definition of $y$).
In the frequency range where the CMB is dominant, the OI 63\,$\mu$m, OI 145\,$\mu$m and CII 158\,$\mu$m transitions drive most of the distortion of the CMB black-body spectrum.
At shorter wavelengths, where the CMB is significantly weaker, other transitions, like FeII 26\,$\mu$m and SiII 35\,$\mu$m, dominate the distortions.

The collisional induced emission is concentrated in overdense regions, where the electron density is able to collisionally modify the fine-structure level populations.
This is consistent with the fact that first (non-linear) star formation events and related feedback processes take place in primordial overdense regions and are effective in ionizing the local medium to a significant degree.
We note that these physical features can be captured thanks to the coupling between the fine-structure emission and the hydro part of the code.
They would be lost with a more naive approach based on simple post-processing estimates.

\begin{figure*}
\centering
\includegraphics[width=18.cm]{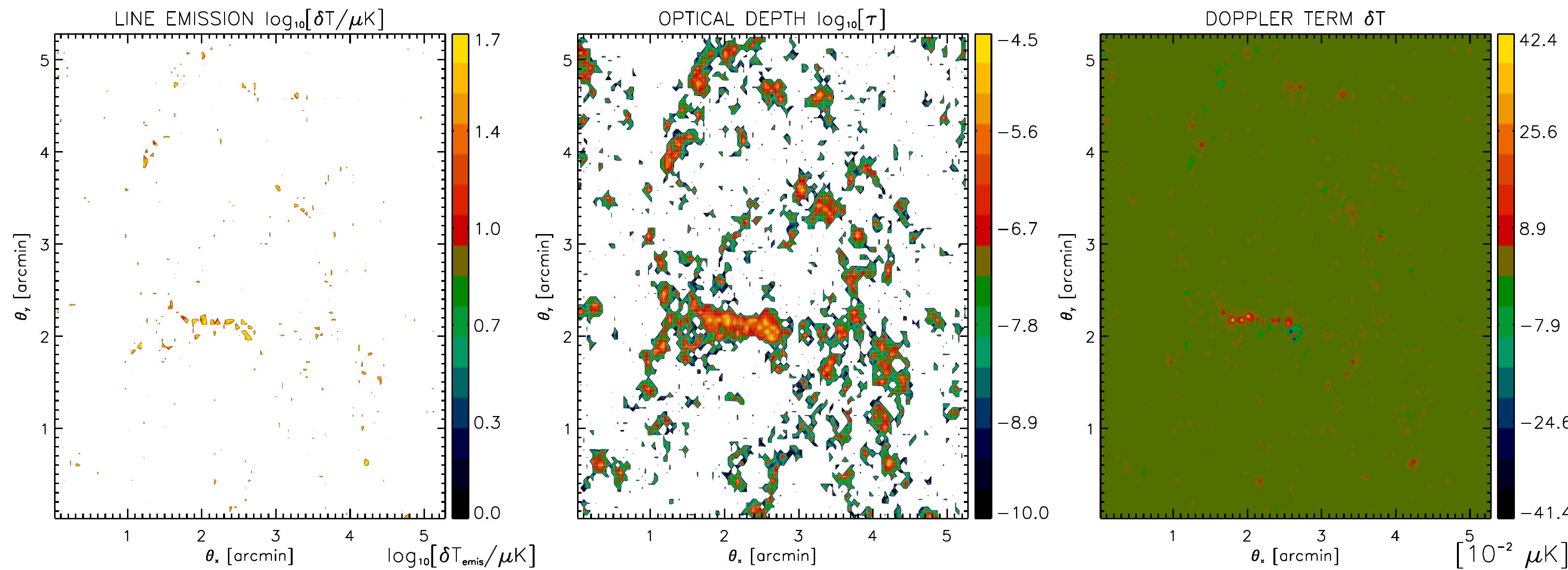}
\caption[fig:maps_CII_z9.5]{
The three panels in this plot refer to the CII 158$\,\mu$m transition as seen at $z=9.5$, which corresponds to an observing frequency of 181$\,$GHz.
{\it Left panel:} Decimal logarithm of the collisionally induced emission measured in
 thermodynamic temperature units [$\mu$K].
{\it Middle panel:} Decimal logarithm of the optical depth associated to the transition. {\it Right panel:} Doppler-induced brightness temperature fluctuations due to the peculiar velocity of the scatterers in units of [$10^{-2}\,\mu$K].
For a square pixel of $\sim 2.5\,$arcsec on a side, a temperature increment of $10^{1.7}\simeq 50\,\mu$K corresponds, at this observing frequency, to a flux of $\sim 3.4\,\mu$Jy.
}
\label{fig:maps_CII_z9.5}
\end{figure*}

To better visualise the impact of metal fine-structure transitions, we derive from the simulated boxes maps of line emission, optical depth and doppler term at different epochs and for different lines.
In particular, in the left panel of Fig.~\ref{fig:maps_CII_z9.5} we find that the CII 158\,$\mu$m emission at $z=9.5$ (which corresponds to $\nu_{\rm obs}=181\,$GHz) can reach a level of almost $\sim 50\,\mu$K 
(in units of thermodynamic temperature at the corresponding resonant frequency) in the highest density cores, where the corresponding optical depth (middle panel) is $\sim 10^{-5}$--$10^{-4}$.
Here we are expressing the emission in units of brightness temperature fluctuations, which relate to specific intensity fluctuations via
\begin{equation}
\frac{\delta T}{T_0} = \frac{\exp{x}-1}{x} \frac{\delta I_{\nu}}{B_{\nu}(T_0)},
\label{eq:dTtodI}
\end{equation}
with $x=h\nu/(k_{\rm B}T_{\rm CMB})$ the adimensional frequency and $T_0$ the CMB temperature monopole at present, $T_0=T_{\rm CMB}[z=0]$. 
This equation assumes that deviations from the CMB black-body intensity law are small.

On the other hand, because CII is widely spread in the IGM, the associated optical depth shows a much smoother distribution in space
($\tau_{\rm CII,\,145}\in$\, [10$^{-8}$,10$^{-5}$] for a $\sim 50\,\%$ of
space), as displayed in the middle panel of Fig.~\ref{fig:maps_CII_z9.5}. 
This is a consequence of metal spreading from stellar evolution due to ongoing star formation and (wind) feedback effects.

Correspondingly, the Doppler effect is not as restricted to the highest density regions as the collisional emission, but more spread to mildly overdense regions. In this case, the amplitude is significantly lower than for the emission term: as the right panel of Fig.~\ref{fig:maps_CII_z9.5} shows in linear scale and independently from the sign flip of the Doppler term (depending on the line-of-sight projected velocity), the amplitude is typically slightly below $0.5\,\mu$K.

During structure growth, more halos hosting collisional emission appear while, at the same time, more metals are spread in the IGM.
Despite the higher metallicity of the medium, the growth of peculiar velocities is modest ($\propto (1+z)^{-1/2}$) and so is the growth of the Doppler term at later epochs.

\begin{figure*}
\centering
\includegraphics[width=18.cm]{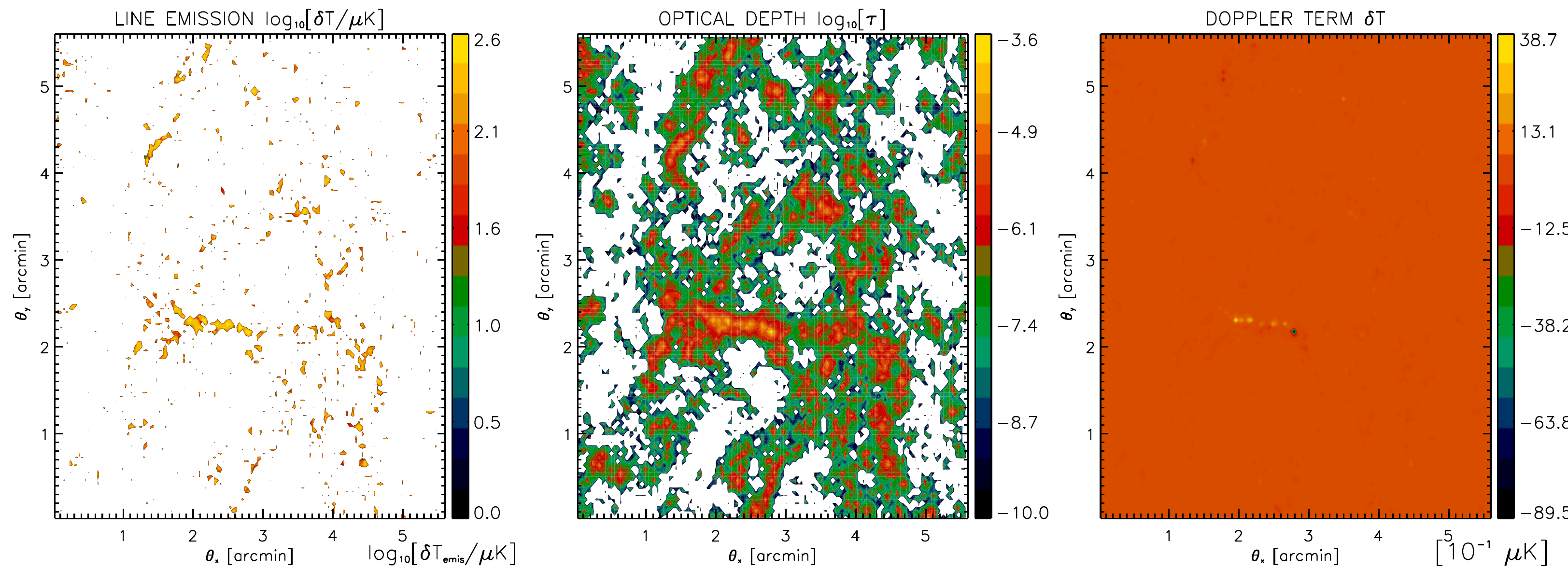}
\caption[fig:maps_CII_z7.5]{
Same as in Fig.~\ref{fig:maps_CII_z9.5}, but referred to the CII 158$\,\mu$m transition as seen at $z=7.5$, corresponding to an observing frequency of $223\,$GHz. 
Differently from Fig.~\ref{fig:maps_CII_z9.5}, results in the right panel are in units of [$10^{-1}\,\mu$K].
For a square pixel of $\sim 2.5\,$arcsec on a side, a temperature increment of $10^{2.6}\simeq 400\,\mu$K corresponds, at this observing frequency, to a flux of $\sim 30\,\mu$Jy.}
\label{fig:maps_CII_z7.5}
\end{figure*}

Fig. \ref{fig:maps_CII_z7.5} is the equivalent of Fig.~\ref{fig:maps_CII_z9.5} at $z=7.5$ ($\nu_{\rm obs}=224\,$GHz) and shows how, while the distribution of metals and associated optical depth are significantly more disperse than at $z=9.5$, the amplitude of the Doppler term has increased mostly due to the larger amount of metals in the IGM, since peculiar velocities have increased by only $\sim 10$\,\% between $z=9.5$ and $z=7.5$. The emission pattern is however significantly different, not only due to the higher amount of clumps where collisions take place, but also to the higher amplitude of the signal in the core of the halos (collisions have not saturated yet at this redshift).

%-----------------------
%-----------------------
 
%-----------------------
%-----------------------

\begin{figure*}
\centering
\includegraphics[width=18.cm]{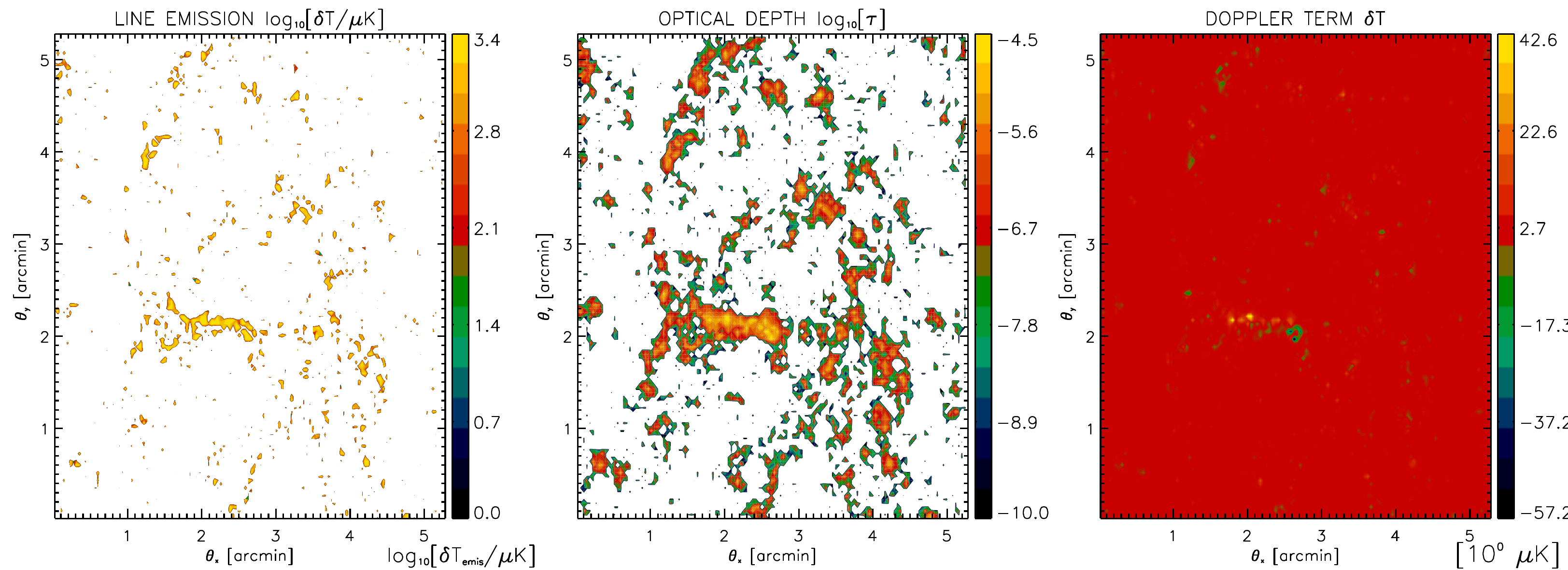}
\caption[fig:maps_OI_z9.5]{Same as in Fig.~\ref{fig:maps_CII_z9.5} but referred to the OI 63$\,\mu$m transition as seen at $z=9.5$, corresponding to an observing frequency of $454\,$GHz. Differently from Fig.~\ref{fig:maps_CII_z9.5}, results in the right panel are in units of [$1\,\mu$K].For a square pixel of $\sim 2.5\,$arcsec on a side, a temperature increment of $10^{3.4}\simeq 2.5\times 10^{3}\,\mu$K corresponds, at this observing frequency, to a flux $\sim 50\,\mu$Jy.}
\label{fig:maps_OI_z9.5}
\end{figure*}

In Fig.~\ref{fig:maps_OI_z9.5} we show the impact of the OI 63$\,\mu$m line at $z=9.5$.
The optical depth is only slightly larger than the CII 158$\,\mu$m transition, while, as can be seen by comparing to Fig.~\ref{fig:maps_CII_z9.5}, the Doppler effect for the OI 63$\,\mu$m transition is about 100 times larger (expressed in units of thermodynamic temperature). This is a combination of the 8 times lower
amplitude of the (background) CMB intensity at the resonant frequency of OI 63$\,\mu$m and of the $\sim 50$ times difference in the factor relating intensity to temperature fluctuations in Eq.~\ref{eq:dTtodI}.
The emission for the OI 63$\,\mu$m case is also significantly higher, and this is caused by the higher oscillation strength of this transition.
In this case, the emission may reach values as high as 10$^3\,\mu$K, which corresponds for a square beam of linear size of $\sim 2.5$\,arcsec, to a flux close to 20$\,\mu$Jy, i.e. about an order of magnitude larger than the corresponding CII 158$\,\mu$m transition.

%%-----------------------
%\begin{figure*}
%\centering
%\includegraphics[width=18.cm]{./Figures/plot_maps_OI_63_z7p5.eps}
%\caption[fig:maps_OI_z7.5]{Same as in Fig.~\ref{fig:maps_CII_z9.5} but referred to the OI 63$\,\mu$m transition as seen at $z=7.5$, corresponding to an observing frequency of $558\,$GHz. }
%\label{fig:maps_OI_z7.5}
%\end{figure*}
%%-----------------------

%-----------------------
%-----------------------

\subsection{Spectral distortions induced on the CMB}

\begin{figure}
\centering
\includegraphics[width=9.cm]{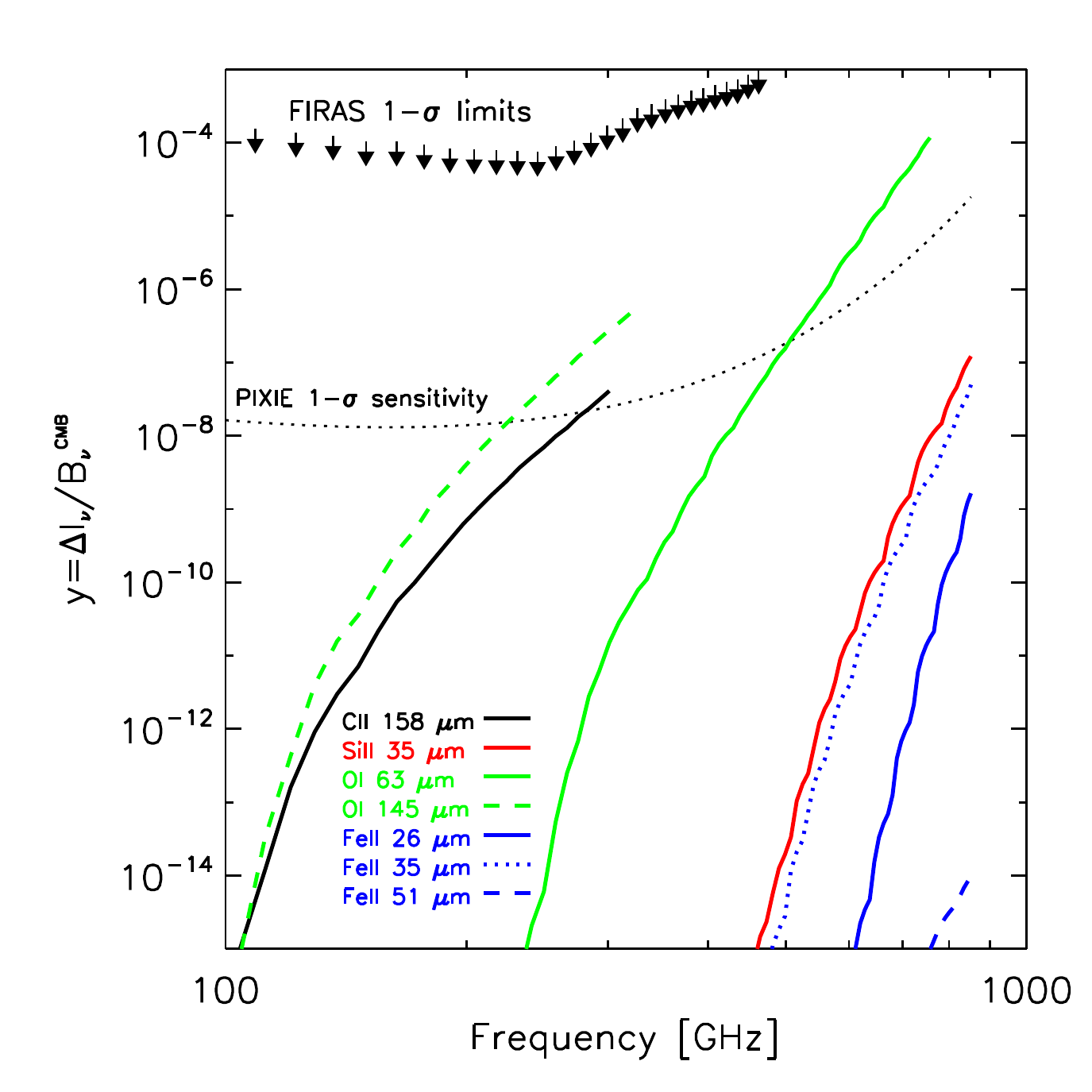}
\caption[fig:ydist]{
Spectral distortion induced by collisionally excited emission on the different resonant transitions considered in this work versus the observing frequency. The upper limits (at 1-$\sigma$) imposed by FIRAS observations \citep[][]{mather94} are displayed by the black arrows in the top part of the figure, while the 1--$\sigma$ sensitivity level foreseen for PIXIE \citep[][]{pixie} is displayed by the black, dotted line. For a fixed resonant redshift, longer wavelength transitions (like CII 158\,$\mu$m or OI 145\,$\mu$m) introduce distortions at the longer wavelengths ($\nu_{\rm obs}\lesssim  200$\,GHz) for which the impact of local dust is less important.
}
\label{fig:ydist}
\end{figure}

COBE/FIRAS \citep[][]{mather94} set strong upper limits to the deviations from a black-body of the measured CMB  spectrum, with a typical RMS deviation at the level of 0.01\,\%.
We next compute the amplitude of the distortion on the CMB black-body spectrum introduced by the cooling lines considered in this paper. While the differential effect generated by each line at each redshift has already been shown in the right panel of Fig.~\ref{fig:fig1-3panels}, in Fig.~\ref{fig:ydist} we show the cumulative\footnote{
By {\it cumulative} we refer to an integration through all resonant redshifts contributing to a given observing frequency.
} effect of all lines considered in this work at different observing frequencies, following the color coding introduced in the right panel of Fig.~\ref{fig:fig1-3panels}. 
The black solid line refers to the CII 158$\,\mu$m transition, which lies just below the OI 145$\,\mu$m line, and both are the dominant contributors for observed frequencies below 300\,GHz.
At $\nu_{\rm obs}\sim 300\,$GHz the distortion induced by these two lines becomes larger than $\sim10^{-8}$, which is about 10$^3$ smaller than current FIRAS upper limits (displayed in Fig.~\ref{fig:ydist} by black arrows in the top of the panel).

We do not show predictions for these two transitions at higher observing frequencies since they correspond to resonant redshifts below $z_{\rm min}=5$, the minimum reached by the simulation.
If however the behaviour of those curves is extrapolated to higher frequencies, then one finds plausible that the distortion due to the two lines would cross the amplitude found for the OI 63$\,\mu$m transition (green solid line) at $\nu_{\rm obs}\sim $600--800$\,$GHz. At these frequencies the  distortion of the CMB induced by OI is $\sim 10^{-4}$. This amplitude should fall well within the range of detectability of future missions like the Primordial Inflation Explorer \citep[PIXIE,][]{pixie}.

The contribution from other species like silicon and iron becomes important only at high frequencies ($\nu_{\rm obs}>800\,$GHz), although they increase very rapidly due to the steep fall in intensity of the CMB in the Wien tail.
Above $\nu_{\rm obs}=300\,$GHz the presence of local and high-$z$ dust and the Cosmic Infrared Background (CIB) generated at $z\in$ [2,5] becomes increasingly important. The challenge lies therefore in detecting the signature of these very early metals embedded in a higher background of emission generated at lower redshifts.
We shall make our first attempts to address this in the following sub-sections.

%-----------------------
%----------------------- 

\subsection{The angular power spectrum induced by metals}

Equation~\ref{eq:RT3} shows that there are {\it a priori} three different terms contributing to the angular anisotropies generated by metals during reionization:
the {\it blurring term}, the {\it Doppler term} and the {\it emission term}.
The blurring of the {\it original} CMB anisotropies is relevant only on very large scales (small $l$-s) and is proportional to the line optical depth.
In the previous section, we have seen that the Doppler term gives rise to {\it new} temperature anisotropies that are significantly smaller than those generated by the emission term.

\begin{figure}
\centering
\includegraphics[width=9.cm]{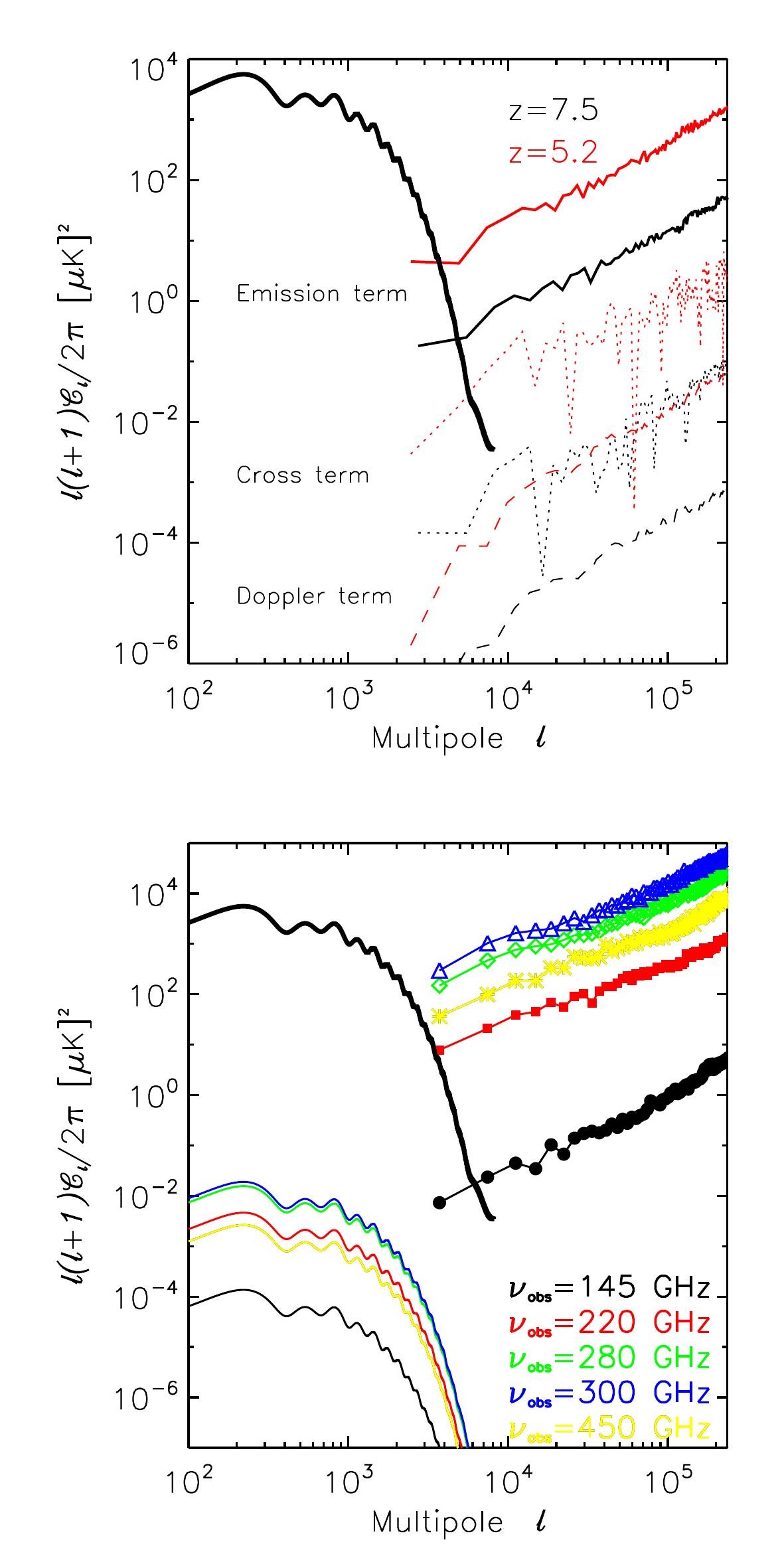}
\caption[fig:angCls]{
{\it Top panel:}
Angular power spectrum of the temperature angular anisotropies induced by the CII 158$\,\mu$m at $z=5.2$ (red color) and $7.5$ (blue color) for observations with $\Delta \nu_{\rm obs}/\nu_{\rm obs}\simeq 5\times 10^{-3}$. We consider separately the contribution from the emission term (solid lines), the Doppler term (dashed lines) and the cross term (dotted lines). The total contribution is practically given by the emission term. In both panels of the figure the intrinsic CMB angular power spectrum consistent with {\it Planck} observations is given by the black, thick solid line.
{\it Bottom panel:}
Cumulative contribution of the collisional emission and Doppler terms from all considered transitions to the angular power spectrum of CMB anisotropies at different observing frequencies. Each observing frequency corresponds to a different color and symbol type (filled black circles for 145\,GHz, filled red squares for 220\,GHz, yellow asterisks for 450\,GHz, green empty diamonds for 280\,GHz and blue empty triangles for 300\,GHz). The thin, coloured solid lines in the bottom left corner of this panel correspond to the (small) contribution of the blurring term at large angular scales \citep[as computed in][]{basu04}.
Note that the prediction for $\nu_{obs}=450\,$GHz misses the contribution from the CII 158$\,\mu$m and OI 145$\,\mu$m transitions since those would be observed at redshifts lower than the minimum output redshift of the simulations.
}
\label{fig:angCls}
\end{figure}

This is also reflected in the top panel of Fig.~\ref{fig:angCls}, where we show the amplitude of the various contributions to the total angular power spectrum for the CII 158$\,\mu$m transition at redshift $z=7.5$ and $z=5.2$, as extracted from the corresponding temperature maps obtained after integrating Eq.~\ref{eq:RT3} in those redshift snapshots and converting intensity fluctuations into temperature fluctuations via Eq.~\ref{eq:dTtodI}. In this way, we can also build maps of temperature fluctuations for the emission and Doppler contributions separately, $\delta T_{\rm emis} $ and $\delta T_{\rm Doppler}$. 
The total amplitude is dominated by (and indistinguishable from) the contribution of the emission term, while the Doppler term is negligible. Since we are plotting power, there exists a cross term coupling emission and the Doppler term (equal to $2 \,\delta T_{\rm emis} \times \delta T_{\rm Doppler}$), which is displayed by the dotted lines.
For the sake of comparison, we also display the intrinsic angular power spectrum of the CMB anisotropies consistent with {\it Planck} measurements \citep[$C_l^{\rm CMB,\,intr}$,][given by the thick black solid line at $ l \lesssim10^4$ in both panels in Fig.~\ref{fig:angCls}.]{planck_parameters_15}.

The bottom panel of the figure reflects the contributions from all lines at given observing frequencies: black, red, green, blue and yellow colours correspond to 145, 220, 280, 300 and 450\,GHz, respectively. These frequencies sample the peak of the CMB black-body spectrum and the Wien tail where the impact of metals is expected to be more relevant.
The symbols joined by the solid lines refer to the combined Doppler and emission terms, while the thin solid lines provide the absolute value of the blurring term, which equals  
$-2 \tau_{\nu_{\rm obs}} C_l^{\rm CMB,\,intr}$ \citep{basu04}.
The notation $\tau_{\nu_{\rm obs}}$ denotes the effective metal optical depth at a given observing frequency and constitutes the sum of all the average optical depths estimated for each transition at the corresponding resonant redshift:
\begin{equation}
\tau_{\nu_{\rm obs}} = \sum_{X,ij} \bar{\tau}^{ij}_{X,\nu}(z_r=\nu^{ij}_X/\nu_{\rm obs}-1).
\label{eq:efftau}
\end{equation}
At relatively low frequencies (at or below 145\,GHz), the corresponding resonant redshifts are typically above $z\sim 15$ for the lines we are considering, and thus their predicted impact on the angular power spectrum is small.
For $\nu_{\rm obs}=145\,$GHz the effect associated to metals intercept the intrinsic CMB angular power spectrum at $l\sim 10^4$, at an amplitude of ${\cal D}_l = l(l+1)C_l/(2\pi) \sim 10^{-2}\,$[$\mu$K]$^2$. However, the signature of metals rapidly increases with frequency (as the effective resonant redshift decreases and samples the epoch of reionization). At $\nu_{\rm obs}=220\,$GHz, the metal-induced signal on small scales is about 100--1000 times larger than at $\nu_{\rm obs}=145\,$GHz, intercepting the intrinsic CMB angular power spectrum at $l\sim 5000$ ($D_l \sim 10\,$[$\mu$K]$^2$). The signal from metals keeps increasing with frequency, since more metals are present at lower redshifts.
Our result for $\nu_{\rm obs}=450\,$GHz has however an amplitude lower than at 280 and 300\,GHz, due to the fact that at this frequency the resonant redshifts for the (dominant) CII 158$\,\mu$m and OI 145$\,\mu$m transitions fall below $z_{\rm min}=5$, the minimum redshift reached by our simulation. 
While the amplitude of the metal-induced signals increases with increasing frequency, the impact of dust emission and the CIB becomes also more important and the need for component separation techniques becomes critical (as it shall be addressed in a subsequent section).

It is also interesting to note that the impact of clustering of metal polluted regions can be seen as the effective resonant redshift decreases: the pattern of the metal induced anisotropies resembles that of isolated, random regions in space at low frequencies ($\nu_{\rm obs} \sim 145\,$GHz) since $C_l \sim {\rm const}$ and $D_l\propto l^2$ (Poissonian), while the slope of $D_l$ becomes shallower at higher frequencies, reflecting the impact of metal clustering at lower redshifts.

We should keep in mind that our approach is approximate in the sense that, while combining contributions from different lines at different redshifts, we are adding the power spectrum for each line at each resonant redshift under a {\em different} effective spectral resolution $\Delta \nu_{\rm obs}/\nu_{\rm obs}$. As described at the end of Sect.~\ref{sec:interac.one}, we are forcing the width of the shell $\Delta s$ to be the simulated box size. According to Eq.~\ref{eq:Ds} this corresponds to a redshift dependent spectral resolution. The value of $\Delta \nu_{\rm obs}/\nu_{\rm obs}$ changes however by a factor $\lesssim 3$ in the entire redshift range under consideration. This, as we discuss below, has a modest impact on the amplitude of $C_l$.

The blurring term can be computed as in \citet[][]{basu04} and, for sake of clarity, it is shown only in the bottom panel of Fig.~\ref{fig:angCls}.
It is worth noting that the low values of the optical depth associated to the transitions make the detection of the blurring term challenging, particularly if one takes into account aspects related to inter-channel calibration and beam characterization uncertainties, as done in e.g. \citet{chm_metals06}. On the positive side, as noted in that work, given an intrinsic CMB anisotropy field, the pattern of the blurring term can be predicted and this should enable signal extraction approaches of a matched-filter type.

%-----------------------
%-----------------------

\subsection{The context provided by current predictions and observations}

\begin{figure}
\centering
\includegraphics[width=9.cm]{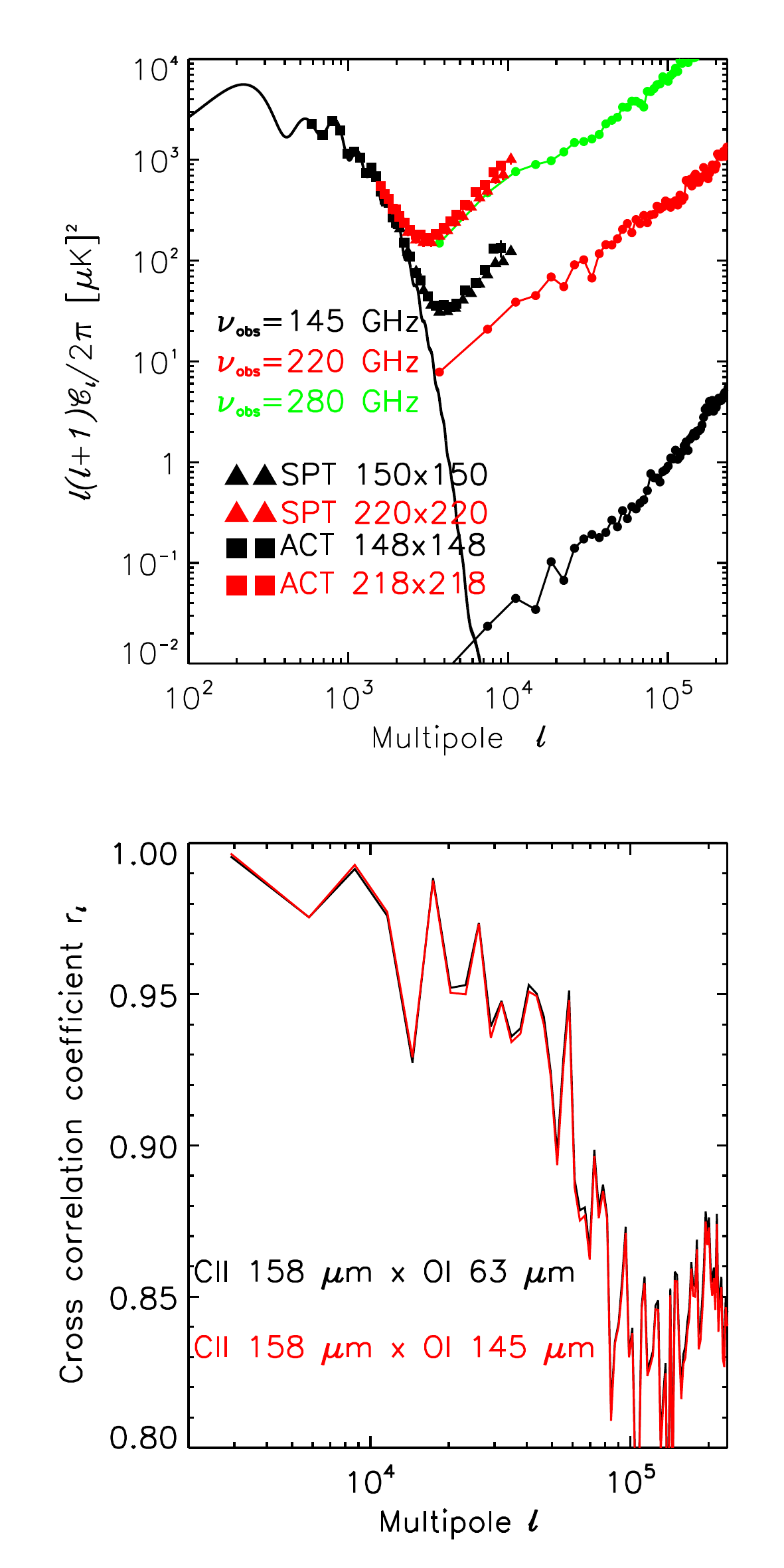}
\caption[fig:crossCls]{
{\it Top panel:} Comparison of the cooling line induced angular power spectrum for frequencies $\nu=145,\,220$\,GHz (displayed by black and red solid lines joined by filled circles) with observations from ACT \citep[filled squares,][]{ACT_LATEST} and SPT \citep[filled triangles,][]{SPT_2015} at similar frequencies. The ratio of the cooling line emission at 145 and 220\,GHz to the total amplitudes measured by ACT and SPT at those same frequencies  increases with frequency. For the sake of completeness the simulation results for $\nu_{\rm obs}=280\,$GHz are also displayed by the green solid line (joined by filled green circles).
{\it Bottom panel:} Cross-correlation coefficient from temperature anisotropies induced by different line pairs located at the same redshift ($z=9.5$): CII 158$\,\mu$m and OI 63$\,\mu$m (black lines) and CII 158$\,\mu$m and OI 145$\,\mu$m (red).
}
\label{fig:crossCls}
\end{figure}

The top panel of Fig.~\ref{fig:crossCls} compares our predictions for the metal-induced signals with state-of-the-art measurements from the Atacama Cosmology Telescope \citep[ACT,][]{ACT_LATEST} and South Pole Telescope \citep[SPT,][]{SPT_2015}  collaborations. Current measurements of CMB intensity anisotropies at 145\,GHz are between 3 and 4 orders of magnitude above our predictions, but only a factor of 10--50 higher at $\nu_{\rm obs}\simeq 220\,$GHz. The signal at this latter frequency and on small angular scales is known to be dominated by dust emission in local and high-$z$ galaxies (which conform the CIB).

As already noted by \citet[][]{pallottini15}, the detection of the CII 158\,$\mu$m signal is challenging. When comparing their predictions for the angular power spectrum associated to this line at $z\simeq 6$ with ours, we find that our estimates fall a factor $\sim 100$ higher in ${\cal D}_l$ amplitude (which corresponds to a factor of $\sim 10$ higher in emission amplitude) for $\Delta \nu_{\rm obs}/\nu_{\rm obs}\sim 5\times 10^{-3}$. At the same time, the amplitude of the ${\cal D}_l$ blurring term in \citet[][]{pallottini15}, for the same redshift, is a factor $\sim 100$ higher than our estimate.
Assuming that the abundance of carbon is similar in both sets of simulations, 
this points to a different  level of efficiency at collisional excitations, thus reflecting the many uncertainties in the modelling of small-scale physics (cooling, metal spreading and mixing, stellar yields, mass and energetics assumed for early SNe, feedback prescriptions, numerical resolution, etc.).

Since different metals probe similar regions in the IGM as they spread through it, the signal generated by different transitions is expected to be spatially correlated. In the bottom panel of Fig.~\ref{fig:crossCls} we compute the correlation coefficient between the total temperature anisotropy maps generated by two different transitions. The correlation coefficient is defined as
\begin{equation}
r_l = \frac{C_l^{X,\,Y}}{\sqrt{C_l^{X,\,X}C_l^{Y,Y}}},
\label{eq:corcoeff}
\end{equation}
where the superscripts $X$ and $Y$ refer to two different transitions.
Our simulations suggest that the correlation at $z=9.5$ between the CII 158\,$\mu$m and the OI 63\,$\mu$m line pairs (or the CII 158\,$\mu$m and  OI 145\,$\mu$m pairs) is high, with cross correlation coefficients typically above 0.85.
The details of the spatial distribution of different metal species on small scales around halos may differ, but on larger scales they prove quite similar.
This opens the possibility of combining observations at different frequencies to extract the weak signal generated by metals:  when targeting a given redshift $z_t$ during reionization, one could combine observations at different frequencies $\nu_{\rm obs,\,1}$ and $\nu_{\rm obs,\,2}$, such that $\nu_{\rm obs,\,1} / \nu_{\rm obs,\,2} = \nu_X / \nu_Y$ and $\nu_{\rm obs,\,1} = \nu_X \, (1+z_t)$, with $\nu_X,\,\nu_Y$ the (rest) resonant frequencies of the transitions under study.
This would alleviate the impact of noise and systematics that do not correlate when using different frequency channels.

%-----------------------
%-----------------------

\subsubsection{Impact of spectral resolution}

\begin{figure}
\centering
\includegraphics[width=9.cm]{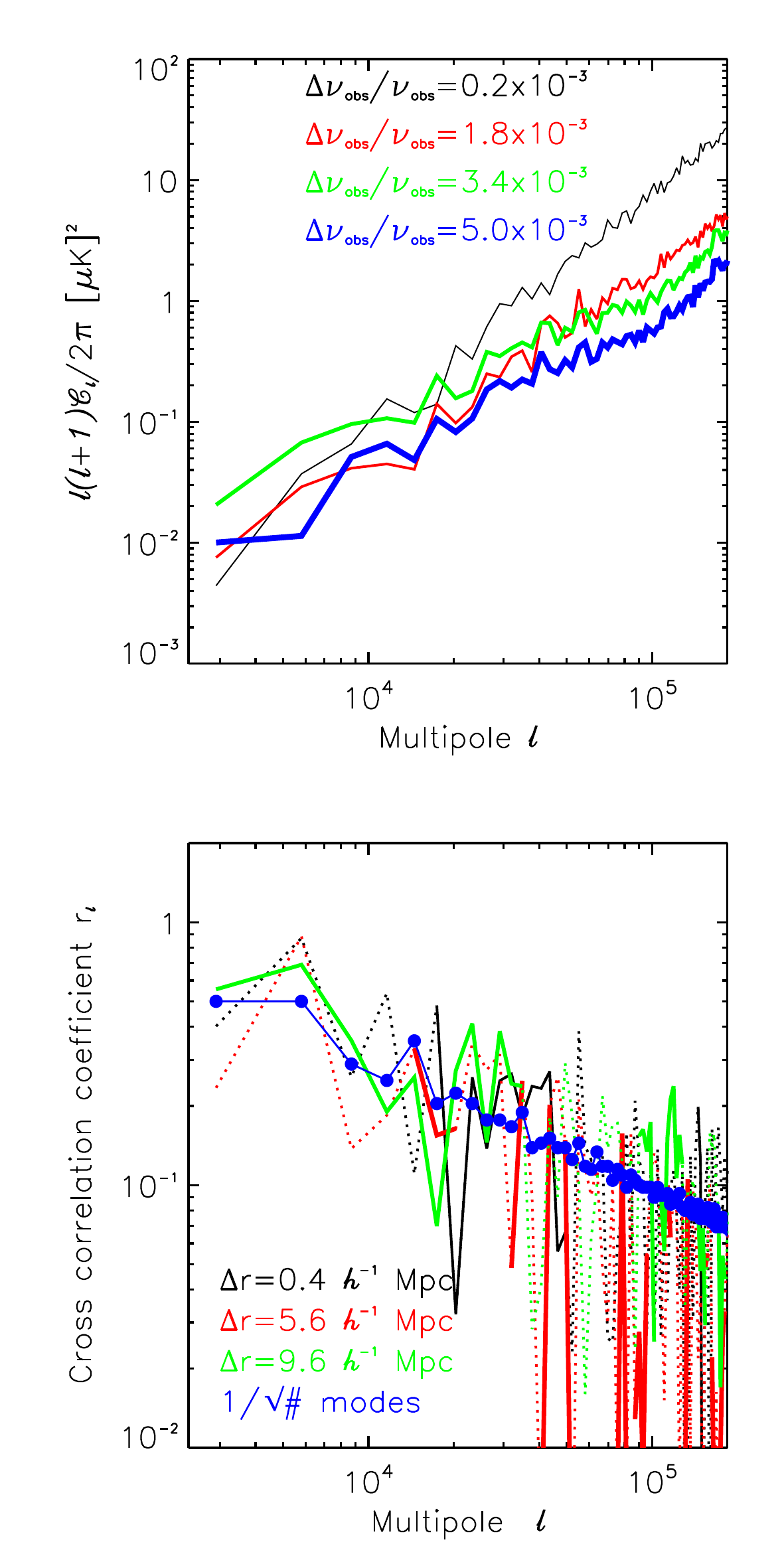}
\caption[fig:deltanuovernu]{
{\it Top panel:} Angular power spectra of the CII 158\,$\mu$m transition from shells centered upon $z=9.5$ and widths resulting from  different choices of spectral resolution: the values $\Delta \nu_{\rm obs}/\nu_{\rm obs}=(0.2, \,1.8,\,3.4,\,5.0)\times 10^{-3}$ are displayed by blue, red, green, and blue lines of increasing thickness, respectively.
{\it Bottom panel:} Cross-correlation coefficient from temperature anisotropies induced by the same line CII 158$\,\mu$m at $z=9.5$ in 
two 
different radial shells of width equal to 400\,$h^{-1}$\,kpc (comoving units). These two shells are taken to lie at a distance $\Delta r$ of 
0.4 (black lines), 
5.6 (red lines), and 
9.6\,$h^{-1}$\,Mpc, (green lines).
Solid (dotted) lines correspond to positive (negative) values of $r_l$, and blue filled circles display the expected scaling for pure noise ($\propto 1/\sqrt{N_k}$, with $N_k$ the number of modes in each radial $k$-bin).
}
\label{fig:deltanuovernu}
\end{figure}

Another interesting characteristic of these signals is their particular dependence on the spectral resolution of the observations. As first noted by \citet[][]{righi08}, when reducing the width of the redshift shell giving rise to the signal we are sensitive to more radial small scale (high-$k$) modes of the anisotropy power spectrum, thus increasing the amplitude of the clustered part of the signal. At the same time, by narrowing the effective redshift window we are also sampling a smaller cosmological volume, and this increases the shot noise contribution to the total measured anisotropy.
In the top panel of Fig.~\ref{fig:deltanuovernu}, we show the angular power spectra for different spectral resolution configurations: for narrower ones the spectrum is merely Poissonian (${\cal D}_l\propto l^2$), since it is dominated by the shot noise contribution, and it increases with decreasing width shell.
On the contrary, for wider shell configurations, the shot noise contribution becomes less important and hints to the clustered contribution on the largest angular scales in the form of a shallower $l$-slope for ${\cal D}_l$.

Since the anisotropy generated in narrow redshift shells is dominated by the shot noise component, the signal from different (but nearby) radial shells should be largely uncorrelated.
In the bottom panel of Fig.~\ref{fig:deltanuovernu} we show the cross-correlation coefficient between shell pairs whose members lie at comoving distances of $\Delta r=$ 0.4, 5.6, and 9.6\,$h^{-1}$\,Mpc. The width of the shells corresponds to $\Delta \nu_{\rm obs}/\nu_{\rm obs}=0.2\times 10^{-3}$. Where positive, lines are solid, and dotted otherwise. Regardless of the inter-shell distance $\Delta r$, the cross correlation coefficient oscillates around zero with a scaling close to $\propto 1/\sqrt{N_k}$, where $N_k$ is the number of 2-D Fourier modes falling within a given radial-$k$ bin. This analysis indicates that the emission in neighbouring redshift shells is uncorrelated and independent from $\Delta r$. This characteristic allows us to remove contamination from foregrounds with a smooth frequency dependence.

%-----------------------
%----------------------- 

\section{Discussion and conclusions}

\label{sec:discon}

Using the hydrodynamical simulations of \citet[][]{Maio2010, Maio2011}, we have estimated the impact of some fine-structure cooling lines (CII 158\,$\mu$m, SiII 35\,$\mu$m, OI 63\,$\mu$m, OI 145\,$\mu$m, FeII 26\,$\mu$m, FeII 36\,$\mu$m and FeII 51\,$\mu$m) on the CMB.
We find that the OI 145\,$\mu$m and CII 158\,$\mu$m lines play a dominant role in the distortions introduced in the CMB black-body spectrum at low frequencies ($\nu < 400$\,GHz), whereas OI 63\,$\mu$m becomes dominant at higher frequencies ($\nu \in [400, 1200]$\,GHz).
In redshift space, the dominant sources of spectral distortions are the OI lines at high redshift ($z>15$), and the FeII 26, FeII 35\,$\mu$m and SiII 35\,$\mu$m transitions at lower redshifts, mostly due to the fact that they show up at very high frequencies ($\nu > 1000$\,GHz), where the CMB intensity has dropped considerably in the Wien region. The amplitude of the distortion induced in the CMB black-body spectrum amounts up to $y=\Delta I_{\nu}/B_{\nu}(T_0)\sim 10^{-8}$--$10^{-6}$ for $\nu < 400$\,GHz, and even higher amplitudes ($\sim 10^{-4}$) at $\nu\simeq 750$\,GHz, still below the upper limits found by FIRAS, but well within the detectability range of PIXIE.

We find that our predictions for CII 158\,$\mu$m tend to be higher than some previous estimates \citep[][]{pallottini15, silva15}.
More specifically, our angular power spectrum amplitude is about a factor of 100 above predictions from \citet[][]{pallottini15}, requiring the amount of collisionally excited CII in that work to be about a factor of 10 lower than in our simulation.
This difference may be partially explained by the consistent inclusion of fine-structure cooling in our modelling and consequent more efficient star formation and metal spreading.
This interpretation is also consistent with their estimate for the blurring term/optical depth $\tau_{CII,\,158\,\mu m}$ being a factor $\sim 100$ higher than in our computations, if we assume that in both works the $z=6$ carbon abundance is similar. However, as we could not estimate their CII metallicity at any redshift, the CII abundance in their work might be very different from ours. 

On the other hand, the work of \citet[][]{silva15} allows for a more straightforward comparison of the spectral distortion induced by the CII 158\,$\mu$m transition. Their Fig.~3 shows a scatter in the redshift dependence of the average CII 158\,$\mu$m specific intensity ($I_{\nu}^0(z)$) of about an order of magnitude among the different models those authors have considered. Our simulation output is typically a factor of 2--3 above their highest amplitude model in the redshift range $z\in [5,8.5]$.

When comparing our clustering amplitude with the predictions from \citet[][]{righi08} we need to re-scale for the different choices of $\Delta \nu_{\rm obs}/\nu_{\rm obs}$. In their Fig.~10 the amplitude of the CII 158\,$\mu$m clustering angular power spectrum ${\cal D}_l$ at $l\simeq 4000$ is roughly 0.3\,($\mu$K)$^2$, and when scaling the spectral resolution from their case ($\Delta \nu_{\rm obs}/\nu_{\rm obs}=0.33$) to ours ($\Delta \nu_{\rm obs}/\nu_{\rm obs}=5\times 10^{-3}$) one should multiply by a factor of $\sim 62$ (see their Fig.~7). At $l= 4000$ their scaled clustering ${\cal D}_l$ amplitude should be close to 20\,($\mu$K)$^2$ at $\nu_{\rm obs}=353$\,GHz. This frequency corresponds to a resonant redshift of $z=4.4$,
%% , lower than the minimum redshift reached by our simulations 
yet, our total amplitude at $z=5.2$ at $l=4000$ is close to 4\,($\mu$K)$^2$, and is mostly dominated by the clustering term (rather than shot noise) on these scales. Assuming a linear scaling with redshift from $z=7.5$ down to $z=4.4$, our projected ${\cal D}_l$ amplitude at $l=4000$ and $z=4.4$ should be three times larger, i.e. about 13\,($\mu$K)$^2$.
This is about twice lower than the scaled estimates from \citet[][]{righi08}.

The model of \citet[][]{righi08} uses a Kennicutt-type relation between the line bolometric luminosity and the star formation rate that is calibrated locally. This approach is also adopted by \citet[][]{silva15} and partially by \citet[][]{pallottini15}, who additionally scale the CII 158\,$\mu$m luminosity as a function of the halo metallicity {\it in post-processing}.
While the fully analytical approach of \citet[][]{righi08} provides the highest CII 158\,$\mu$m anisotropy estimate, the 'hybrid' \citet[][]{pallottini15} and \citet[][]{silva15} predictions, mixing numerical simulations and post-processing scaling laws for emission line luminosities, provide lower estimates. Our predictions based exclusively on numerical simulations (including consistently cooling from the considered fine-structure lines) lie within an intermediate range.
%  On the other hand, our simulations,
By tracking all the occupation levels of the transitions under analysis, we predict amplitudes for the angular power spectra of the blurring term to lie in the range $10^{-6}$--$10^{-2}\,\mu$K. Thus, their detection will require very demanding levels of cross-channel intercalibration and foreground removal.

However, even results from full numerical simulations suffer from uncertainties. In our modelling, stellar lifetimes dictate the temporal evolution of cosmic metal spreading and the occurrence of the heavy elements. Dependencies of stellar lifetimes on metallicity are not very strong and are quite well known. Variations could arise, though, from different prescriptions for low-mass star evolution, binary systems and resulting SNIa rates, which are nevertheless expected to have a minor impact on our results.
Stellar yields might play a role in the amount of metals expelled. High-redshift metal enrichment is led mostly by oxygen and carbon, whose yields are safely constrained and uncertainties should not affect gas cooling and enrichment.
Modelling of the early stellar populations is rather uncertain due our large ignorance on the mass scales of primordial stars (i.e. their IMF).
We have modelled metal spreading by smoothing metallicities over neighbouring particles in order to mimic diffusion of heavy elements.
This is a simple approximation that accounts for such process, however it could lack the necessary details at very small scales. In general, this is a very complex issue and, despite some attempts, a fully satisfying treatment is still missing.

Cosmic gas collapse, structure growth and subsequent metal enrichment are also affected by the background cosmological model. Here we perform all the calculations in a standard $\Lambda$CDM model with a Gaussian input power spectrum, although effects on baryon and metal history of different scenarios, at $z>5$, are possible. We checked that including e.g.
early dark energy \citep[][]{Maio2006},
non-Gaussianities \citep[][]{Maio2011ng, MK2012},
warm dark matter \citep[][]{MV2015},
as well as primordial gaseous flows \citep[][]{Maio2011vb} reflects in changes in the star formation at very early times and only for the collapse of the earliest gas clouds.
The late-time evolution of gas and heavy elements in the different models tend to converge already within the first Gyr.
A full detailed radiative transfer of the individual lines presented above has not been adopted, because we find that \cite[see e.g.][]{PM2012, Maio2016}, with the exception of extreme cases, thermodynamical trends, and hence the resulting level populations, are not significantly altered.

The collisional emission on carbon and oxygen atoms studied here has so far neglected the contribution from the UV pumping Field-Wouthuysen effect presented in \citet[][]{chm_OI_I} and \citet[][]{chm_OI_II,kawasaki12}. This would require a careful evaluation of the UV flux and spectrum around luminous sources. This effect would add on top of the collision-induced distortions computed in this work and would still be associated to specific fine-structure transitions of oxygen and carbon.

As proven in the bottom panel of Fig.~\ref{fig:deltanuovernu}, line-induced signals are largely uncorrelated in neighbouring frequency channels. This constitutes a distinct feature that, in H 21\,cm science, is expected to allow the removal of smoothly frequency varying foregrounds of amplitude much higher than that of the signal of interest \citep[see, e.g.,][]{santos05, chapman12}. While in principle this approach could work also in this context, the confusion arises with the presence of other lines of longer wavelengths coming from redshifts lower than the target line. In this sense, fine-structure lines associated to nitrogen\footnote{
Unfortunately, literature works do not provide reliable estimates for stellar N yields, hence its cosmological evolution could be only vaguely assessed. In this respect, see discussions in e.g. \cite{MT2015}.
}
 not considered here (like NII 121\,$\mu$m or NII 205\,$\mu$m) would still be probing a redshift range similar to that of the CII 158\,$\mu$m and the OI 145\,$\mu$m lines that are shown to contribute at $\nu<300$\,GHz. (see Fig.~\ref{fig:ydist}). Indeed, the contribution from different lines may be unveiled via cross-correlations of multi-frequency observations probing different species at the same redshift, provided a large degree of spatial correlation of the species (as shown in the bottom panel of Fig.~\ref{fig:crossCls}). The amplitude of distortions induced by these lines is steadily growing with frequency as lines probe more recent epochs of the reionization era. This constitutes a very different frequency pattern than the one expected from Comptonization of the CMB spectrum by hot electrons in halos and the IGM during and after reionization \citep[i.e., the $y$-type distortion induced by the thermal Sunyaev-Zeldovich effect,][]{tSZ}. 

Component separation algorithms will thus play a crucial role in the interpretation of multiple, subtle signals distorting the CMB black-body spectrum at different cosmological epochs. The frequency dependence of some of those signals can be well established theoretically \citep[i.e., the $y$-type distortion induced by the Compton scattering of CMB photons off hot electrons, or the $\mu$-type of distortion that may be introduced at very early cosmological times, although there is a wealth of ``intermediate'' cases, see][for reviews]{chluba_sunyaev12, sunyaev_khatri14, chluba_review16}. For other signals associated to atoms and molecules, the frequency dependence is determined by their actual redshift distribution. In this complex landscape, accessing and isolating signals generated during the epoch of reionization requires profiting from the knowledge of their angular and spectral patterns and of their correlation properties in those two domains (angle and frequency). In this context, existing and upcoming experiments of different nature and purpose (like FIRAS, {\it Planck}, ACT, SPT, ALMA, PIXIE, or {\it COrE}) may nicely complement each other.

\begin{acknowledgements}
C.~H.-M. acknowledges the financial support from the Spanish MINECO through project AYA-2015-66211-C2-2-P.
U. M.'s research has received funding from the European Union Seventh Framework Programme (FP7/2007-2013) under grant agreement n. 267251.
We acknowledge the support of the Rechenzentrum Garching. Computations were performed on the machines at the computing centre of the Max Planck Society with CPU time assigned to the Max Planck Institute for Astrophysics.
We made use of the tools offered by the NASA Astrophysics Data Systems for bibliographic research.
\end{acknowledgements}

% Bibtex style:------------
\bibliographystyle{aa}
\bibliography{paper_metals}

\begin{thebibliography}{67}
\expandafter\ifx\csname natexlab\endcsname\relax\def\natexlab#1{#1}\fi

\bibitem[{{Anninos} {et~al.}(1997){Anninos}, {Zhang}, {Abel}, \&
  {Norman}}]{Anninos1997}
{Anninos}, P., {Zhang}, Y., {Abel}, T., \& {Norman}, M.~L. 1997, \na, 2, 209

\bibitem[{{Asplund} {et~al.}(2009){Asplund}, {Grevesse}, {Sauval}, \&
  {Scott}}]{asplund09}
{Asplund}, M., {Grevesse}, N., {Sauval}, A.~J., \& {Scott}, P. 2009, \araa, 47,
  481

\bibitem[{{Barnes} \& {Hut}(1986)}]{barnesandhut86}
{Barnes}, J. \& {Hut}, P. 1986, \nat, 324, 446

\bibitem[{{Basu} {et~al.}(2004){Basu}, {Hern{\'a}ndez-Monteagudo}, \&
  {Sunyaev}}]{basu04}
{Basu}, K., {Hern{\'a}ndez-Monteagudo}, C., \& {Sunyaev}, R.~A. 2004, \aap,
  416, 447

\bibitem[{{Bowman} {et~al.}(2013){Bowman}, {Cairns}, {Kaplan}, {Murphy},
  {Oberoi}, {Staveley-Smith}, {Arcus}, {Barnes}, {Bernardi}, {Briggs}, {Brown},
  {Bunton}, {Burgasser}, {Cappallo}, {Chatterjee}, {Corey}, {Coster},
  {Deshpande}, {deSouza}, {Emrich}, {Erickson}, {Goeke}, {Gaensler},
  {Greenhill}, {Harvey-Smith}, {Hazelton}, {Herne}, {Hewitt},
  {Johnston-Hollitt}, {Kasper}, {Kincaid}, {Koenig}, {Kratzenberg}, {Lonsdale},
  {Lynch}, {Matthews}, {McWhirter}, {Mitchell}, {Morales}, {Morgan}, {Ord},
  {Pathikulangara}, {Prabu}, {Remillard}, {Robishaw}, {Rogers}, {Roshi},
  {Salah}, {Sault}, {Shankar}, {Srivani}, {Stevens}, {Subrahmanyan}, {Tingay},
  {Wayth}, {Waterson}, {Webster}, {Whitney}, {Williams}, {Williams}, \&
  {Wyithe}}]{mwa}
{Bowman}, J.~D., {Cairns}, I., {Kaplan}, D.~L., {et~al.} 2013, \pasa, 30, e031

\bibitem[{{Bromm} \& {Loeb}(2003)}]{Bromm2003}
{Bromm}, V. \& {Loeb}, A. 2003, \nat, 425, 812

\bibitem[{{Cen} \& {Ostriker}(1993)}]{cenostriker93}
{Cen}, R. \& {Ostriker}, J.~P. 1993, \apj, 417, 404

\bibitem[{{Cen} {et~al.}(1990){Cen}, {Ostriker}, {Jameson}, \&
  {Liu}}]{cenostriker90}
{Cen}, R.~Y., {Ostriker}, J.~P., {Jameson}, A., \& {Liu}, F. 1990, \apjl, 362,
  L41

\bibitem[{{Chapman} {et~al.}(2012){Chapman}, {Abdalla}, {Harker}, {Jeli{\'c}},
  {Labropoulos}, {Zaroubi}, {Brentjens}, {de Bruyn}, \& {Koopmans}}]{chapman12}
{Chapman}, E., {Abdalla}, F.~B., {Harker}, G., {et~al.} 2012, \mnras, 423, 2518

\bibitem[{{Chluba}(2016{\natexlab{a}})}]{jens_review_16}
{Chluba}, J. 2016{\natexlab{a}}, \mnras, 460, 227

\bibitem[{{Chluba}(2016{\natexlab{b}})}]{chluba_review16}
{Chluba}, J. 2016{\natexlab{b}}, \mnras, 460, 227

\bibitem[{{Chluba} \& {Sunyaev}(2012)}]{chluba_sunyaev12}
{Chluba}, J. \& {Sunyaev}, R.~A. 2012, \mnras, 419, 1294

\bibitem[{{Das} {et~al.}(2014){Das}, {Louis}, {Nolta}, {Addison},
  {Battistelli}, {Bond}, {Calabrese}, {Crichton}, {Devlin}, {Dicker},
  {Dunkley}, {D{\"u}nner}, {Fowler}, {Gralla}, {Hajian}, {Halpern},
  {Hasselfield}, {Hilton}, {Hincks}, {Hlozek}, {Huffenberger}, {Hughes},
  {Irwin}, {Kosowsky}, {Lupton}, {Marriage}, {Marsden}, {Menanteau}, {Moodley},
  {Niemack}, {Page}, {Partridge}, {Reese}, {Schmitt}, {Sehgal}, {Sherwin},
  {Sievers}, {Spergel}, {Staggs}, {Swetz}, {Switzer}, {Thornton}, {Trac}, \&
  {Wollack}}]{ACT_LATEST}
{Das}, S., {Louis}, T., {Nolta}, M.~R., {et~al.} 2014, \jcap, 4, 014

\bibitem[{{Field}(1958)}]{field58}
{Field}, G.~B. 1958, Proceedings of the IRE, 46, 240

\bibitem[{{George} {et~al.}(2015){George}, {Reichardt}, {Aird}, {Benson},
  {Bleem}, {Carlstrom}, {Chang}, {Cho}, {Crawford}, {Crites}, {de Haan},
  {Dobbs}, {Dudley}, {Halverson}, {Harrington}, {Holder}, {Holzapfel}, {Hou},
  {Hrubes}, {Keisler}, {Knox}, {Lee}, {Leitch}, {Lueker}, {Luong-Van},
  {McMahon}, {Mehl}, {Meyer}, {Millea}, {Mocanu}, {Mohr}, {Montroy}, {Padin},
  {Plagge}, {Pryke}, {Ruhl}, {Schaffer}, {Shaw}, {Shirokoff}, {Spieler},
  {Staniszewski}, {Stark}, {Story}, {van Engelen}, {Vanderlinde}, {Vieira},
  {Williamson}, \& {Zahn}}]{SPT_2015}
{George}, E.~M., {Reichardt}, C.~L., {Aird}, K.~A., {et~al.} 2015, \apj, 799,
  177

\bibitem[{{Gnedin}(1996{\natexlab{a}})}]{gnedin96I}
{Gnedin}, N.~Y. 1996{\natexlab{a}}, \apj, 456, 1

\bibitem[{{Gnedin}(1996{\natexlab{b}})}]{gnedin96II}
{Gnedin}, N.~Y. 1996{\natexlab{b}}, \apj, 456, 34

\bibitem[{{Gnedin} \& {Bertschinger}(1996)}]{gnedinbertschinger96}
{Gnedin}, N.~Y. \& {Bertschinger}, E. 1996, \apj, 470, 115

\bibitem[{{Hern{\'a}ndez-Monteagudo}
  {et~al.}(2007{\natexlab{a}}){Hern{\'a}ndez-Monteagudo}, {Haiman}, {Jimenez},
  \& {Verde}}]{chm_OI_I}
{Hern{\'a}ndez-Monteagudo}, C., {Haiman}, Z., {Jimenez}, R., \& {Verde}, L.
  2007{\natexlab{a}}, \apjl, 660, L85

\bibitem[{{Hern{\'a}ndez-Monteagudo} {et~al.}(2008){Hern{\'a}ndez-Monteagudo},
  {Haiman}, {Verde}, \& {Jimenez}}]{chm_OI_II}
{Hern{\'a}ndez-Monteagudo}, C., {Haiman}, Z., {Verde}, L., \& {Jimenez}, R.
  2008, \apj, 672, 33

\bibitem[{{Hern{\'a}ndez-Monteagudo}
  {et~al.}(2007{\natexlab{b}}){Hern{\'a}ndez-Monteagudo},
  {Rubi{\~n}o-Mart{\'{\i}}n}, \& {Sunyaev}}]{chm_pol_07}
{Hern{\'a}ndez-Monteagudo}, C., {Rubi{\~n}o-Mart{\'{\i}}n}, J.~A., \&
  {Sunyaev}, R.~A. 2007{\natexlab{b}}, \mnras, 380, 1656

\bibitem[{{Hern{\'a}ndez-Monteagudo} {et~al.}(2006){Hern{\'a}ndez-Monteagudo},
  {Verde}, \& {Jimenez}}]{chm_metals06}
{Hern{\'a}ndez-Monteagudo}, C., {Verde}, L., \& {Jimenez}, R. 2006, \apj, 653,
  1

\bibitem[{{Hill} {et~al.}(2015){Hill}, {Battaglia}, {Chluba}, {Ferraro},
  {Schaan}, \& {Spergel}}]{Hill2015}
{Hill}, J.~C., {Battaglia}, N., {Chluba}, J., {et~al.} 2015, Physical Review
  Letters, 115, 261301

\bibitem[{{Hockney} \& {Eastwood}(1988)}]{hockneyandeastwood88}
{Hockney}, R.~W. \& {Eastwood}, J.~W. 1988, {Computer simulation using
  particles}

\bibitem[{{Holmberg}(1941)}]{holmberg41}
{Holmberg}, E. 1941, \apj, 94, 385

\bibitem[{{Kogut} {et~al.}(2016){Kogut}, {Chluba}, {Fixsen}, {Meyer}, \&
  {Spergel}}]{pixie}
{Kogut}, A., {Chluba}, J., {Fixsen}, D.~J., {Meyer}, S., \& {Spergel}, D. 2016,
  in \procspie, Vol. 9904, Space Telescopes and Instrumentation 2016: Optical,
  Infrared, and Millimeter Wave, 99040W

\bibitem[{{Koopmans} {et~al.}(2015){Koopmans}, {Pritchard}, {Mellema},
  {Aguirre}, {Ahn}, {Barkana}, {van Bemmel}, {Bernardi}, {Bonaldi}, {Briggs},
  {de Bruyn}, {Chang}, {Chapman}, {Chen}, {Ciardi}, {Dayal}, {Ferrara},
  {Fialkov}, {Fiore}, {Ichiki}, {Illiev}, {Inoue}, {Jelic}, {Jones}, {Lazio},
  {Maio}, {Majumdar}, {Mack}, {Mesinger}, {Morales}, {Parsons}, {Pen},
  {Santos}, {Schneider}, {Semelin}, {de Souza}, {Subrahmanyan}, {Takeuchi},
  {Vedantham}, {Wagg}, {Webster}, {Wyithe}, {Datta}, \& {Trott}}]{ska}
{Koopmans}, L., {Pritchard}, J., {Mellema}, G., {et~al.} 2015, Advancing
  Astrophysics with the Square Kilometre Array (AASKA14), 1

\bibitem[{{Kusakabe} \& {Kawasaki}(2012)}]{kawasaki12}
{Kusakabe}, M. \& {Kawasaki}, M. 2012, \mnras, 419, 873

\bibitem[{{Maio} {et~al.}(2010){Maio}, {Ciardi}, {Dolag}, {Tornatore}, \&
  {Khochfar}}]{Maio2010}
{Maio}, U., {Ciardi}, B., {Dolag}, K., {Tornatore}, L., \& {Khochfar}, S. 2010,
  \mnras, 407, 1003

\bibitem[{{Maio} {et~al.}(2009){Maio}, {Ciardi}, {Yoshida}, {Dolag}, \&
  {Tornatore}}]{Maio2009}
{Maio}, U., {Ciardi}, B., {Yoshida}, N., {Dolag}, K., \& {Tornatore}, L. 2009,
  \aap, 503, 25

\bibitem[{{Maio} {et~al.}(2007){Maio}, {Dolag}, {Ciardi}, \&
  {Tornatore}}]{maio07}
{Maio}, U., {Dolag}, K., {Ciardi}, B., \& {Tornatore}, L. 2007, \mnras, 379,
  963

\bibitem[{{Maio} {et~al.}(2006){Maio}, {Dolag}, {Meneghetti}, {Moscardini},
  {Yoshida}, {Baccigalupi}, {Bartelmann}, \& {Perrotta}}]{Maio2006}
{Maio}, U., {Dolag}, K., {Meneghetti}, M., {et~al.} 2006, \mnras, 373, 869

\bibitem[{{Maio} {et~al.}(2013){Maio}, {Dotti}, {Petkova}, {Perego}, \&
  {Volonteri}}]{Maio2013}
{Maio}, U., {Dotti}, M., {Petkova}, M., {Perego}, A., \& {Volonteri}, M. 2013,
  \apj, 767, 37

\bibitem[{{Maio} \& {Iannuzzi}(2011)}]{Maio2011ng}
{Maio}, U. \& {Iannuzzi}, F. 2011, \mnras, 415, 3021

\bibitem[{{Maio} \& {Khochfar}(2012)}]{MK2012}
{Maio}, U. \& {Khochfar}, S. 2012, \mnras, 421, 1113

\bibitem[{{Maio} {et~al.}(2011{\natexlab{a}}){Maio}, {Khochfar}, {Johnson}, \&
  {Ciardi}}]{Maio2011}
{Maio}, U., {Khochfar}, S., {Johnson}, J.~L., \& {Ciardi}, B.
  2011{\natexlab{a}}, \mnras, 414, 1145

\bibitem[{{Maio} {et~al.}(2011{\natexlab{b}}){Maio}, {Koopmans}, \&
  {Ciardi}}]{Maio2011vb}
{Maio}, U., {Koopmans}, L.~V.~E., \& {Ciardi}, B. 2011{\natexlab{b}}, \mnras,
  412, L40

\bibitem[{{Maio} {et~al.}(2016){Maio}, {Petkova}, {De Lucia}, \&
  {Borgani}}]{Maio2016}
{Maio}, U., {Petkova}, M., {De Lucia}, G., \& {Borgani}, S. 2016, \mnras, 460,
  3733

\bibitem[{{Maio} \& {Tescari}(2015)}]{MT2015}
{Maio}, U. \& {Tescari}, E. 2015, \mnras, 453, 3798

\bibitem[{{Maio} \& {Viel}(2015)}]{MV2015}
{Maio}, U. \& {Viel}, M. 2015, \mnras, 446, 2760

\bibitem[{{Mather} {et~al.}(1994){Mather}, {Cheng}, {Cottingham}, {Eplee},
  {Fixsen}, {Hewagama}, {Isaacman}, {Jensen}, {Meyer}, {Noerdlinger}, {Read},
  {Rosen}, {Shafer}, {Wright}, {Bennett}, {Boggess}, {Hauser}, {Kelsall},
  {Moseley}, {Silverberg}, {Smoot}, {Weiss}, \& {Wilkinson}}]{mather94}
{Mather}, J.~C., {Cheng}, E.~S., {Cottingham}, D.~A., {et~al.} 1994, \apj, 420,
  439

\bibitem[{{Matteucci} \& {Greggio}(1986)}]{MatteucciGreggio1986}
{Matteucci}, F. \& {Greggio}, L. 1986, \aap, 154, 279

\bibitem[{{Pallottini} {et~al.}(2015){Pallottini}, {Gallerani}, {Ferrara},
  {Yue}, {Vallini}, {Maiolino}, \& {Feruglio}}]{pallottini15}
{Pallottini}, A., {Gallerani}, S., {Ferrara}, A., {et~al.} 2015, \mnras, 453,
  1898

\bibitem[{{Petkova} \& {Maio}(2012)}]{PM2012}
{Petkova}, M. \& {Maio}, U. 2012, \mnras, 422, 3067

\bibitem[{{Planck Collaboration} {et~al.}(2016){Planck Collaboration}, {Ade},
  {Aghanim}, {Arnaud}, {Ashdown}, {Aumont}, {Baccigalupi}, {Banday},
  {Barreiro}, {Bartlett}, \& et~al.}]{planck_parameters_15}
{Planck Collaboration}, {Ade}, P.~A.~R., {Aghanim}, N., {et~al.} 2016, \aap,
  594, A13

\bibitem[{{Pritchard} \& {Loeb}(2012)}]{pritchard12}
{Pritchard}, J.~R. \& {Loeb}, A. 2012, Reports on Progress in Physics, 75,
  086901

\bibitem[{{Renzini} \& {Buzzoni}(1986)}]{RenziniBuzzoni1986}
{Renzini}, A. \& {Buzzoni}, A. 1986, in Astrophysics and Space Science Library,
  Vol. 122, Spectral Evolution of Galaxies, ed. C.~{Chiosi} \& A.~{Renzini},
  195--231

\bibitem[{{Righi} {et~al.}(2008){Righi}, {Hern{\'a}ndez-Monteagudo}, \&
  {Sunyaev}}]{righi08}
{Righi}, M., {Hern{\'a}ndez-Monteagudo}, C., \& {Sunyaev}, R.~A. 2008, \aap,
  489, 489

\bibitem[{{Santos} {et~al.}(2005){Santos}, {Cooray}, \& {Knox}}]{santos05}
{Santos}, M.~G., {Cooray}, A., \& {Knox}, L. 2005, \apj, 625, 575

\bibitem[{{Schneider} {et~al.}(2006){Schneider}, {Omukai}, {Inoue}, \&
  {Ferrara}}]{Schneider2006}
{Schneider}, R., {Omukai}, K., {Inoue}, A.~K., \& {Ferrara}, A. 2006, \mnras,
  369, 1437

\bibitem[{{Shimasaku} {et~al.}(2006){Shimasaku}, {Kashikawa}, {Doi}, {Ly},
  {Malkan}, {Matsuda}, {Ouchi}, {Hayashino}, {Iye}, {Motohara}, {Murayama},
  {Nagao}, {Ohta}, {Okamura}, {Sasaki}, {Shioya}, \& {Taniguchi}}]{subaruLAEs}
{Shimasaku}, K., {Kashikawa}, N., {Doi}, M., {et~al.} 2006, \pasj, 58, 313

\bibitem[{{Silva} {et~al.}(2015){Silva}, {Santos}, {Cooray}, \&
  {Gong}}]{silva15}
{Silva}, M., {Santos}, M.~G., {Cooray}, A., \& {Gong}, Y. 2015, \apj, 806, 209

\bibitem[{{Springel}(2005)}]{gadget}
{Springel}, V. 2005, \mnras, 364, 1105

\bibitem[{{Suginohara} {et~al.}(1999){Suginohara}, {Suginohara}, \&
  {Spergel}}]{suginohara2spergel99}
{Suginohara}, M., {Suginohara}, T., \& {Spergel}, D.~N. 1999, \apj, 512, 547

\bibitem[{{Sunyaev} \& {Khatri}(2013)}]{sunyaev_khatri14}
{Sunyaev}, R.~A. \& {Khatri}, R. 2013, International Journal of Modern Physics
  D, 22, 1330014

\bibitem[{{Sunyaev} \& {Zeldovich}(1972)}]{tSZ}
{Sunyaev}, R.~A. \& {Zeldovich}, Y.~B. 1972, Comments on Astrophysics and Space
  Physics, 4, 173

\bibitem[{{Thielemann} {et~al.}(2003){Thielemann}, {Argast}, {Brachwitz},
  {Hix}, {H{\"o}flich}, {Liebend{\"o}rfer}, {Martinez-Pinedo}, {Mezzacappa},
  {Panov}, \& {Rauscher}}]{Thielemann2003}
{Thielemann}, F.-K., {Argast}, D., {Brachwitz}, F., {et~al.} 2003, Nuclear
  Physics A, 718, 139

\bibitem[{{Tornatore} {et~al.}(2010){Tornatore}, {Borgani}, {Viel}, \&
  {Springel}}]{Tornatore2010}
{Tornatore}, L., {Borgani}, S., {Viel}, M., \& {Springel}, V. 2010, \mnras,
  402, 1911

\bibitem[{{van den Hoek} \& {Groenewegen}(1997)}]{vdHoek1997}
{van den Hoek}, L.~B. \& {Groenewegen}, M.~A.~T. 1997, \aaps, 123

\bibitem[{{van Haarlem} {et~al.}(2013){van Haarlem}, {Wise}, {Gunst}, {Heald},
  {McKean}, {Hessels}, {de Bruyn}, {Nijboer}, {Swinbank}, {Fallows},
  {Brentjens}, {Nelles}, {Beck}, {Falcke}, {Fender}, {H{\"o}randel},
  {Koopmans}, {Mann}, {Miley}, {R{\"o}ttgering}, {Stappers}, {Wijers},
  {Zaroubi}, {van den Akker}, {Alexov}, {Anderson}, {Anderson}, {van Ardenne},
  {Arts}, {Asgekar}, {Avruch}, {Batejat}, {B{\"a}hren}, {Bell}, {Bell}, {van
  Bemmel}, {Bennema}, {Bentum}, {Bernardi}, {Best}, {B{\^i}rzan}, {Bonafede},
  {Boonstra}, {Braun}, {Bregman}, {Breitling}, {van de Brink}, {Broderick},
  {Broekema}, {Brouw}, {Br{\"u}ggen}, {Butcher}, {van Cappellen}, {Ciardi},
  {Coenen}, {Conway}, {Coolen}, {Corstanje}, {Damstra}, {Davies}, {Deller},
  {Dettmar}, {van Diepen}, {Dijkstra}, {Donker}, {Doorduin}, {Dromer}, {Drost},
  {van Duin}, {Eisl{\"o}ffel}, {van Enst}, {Ferrari}, {Frieswijk}, {Gankema},
  {Garrett}, {de Gasperin}, {Gerbers}, {de Geus}, {Grie{\ss}meier}, {Grit},
  {Gruppen}, {Hamaker}, {Hassall}, {Hoeft}, {Holties}, {Horneffer}, {van der
  Horst}, {van Houwelingen}, {Huijgen}, {Iacobelli}, {Intema}, {Jackson},
  {Jelic}, {de Jong}, {Juette}, {Kant}, {Karastergiou}, {Koers}, {Kollen},
  {Kondratiev}, {Kooistra}, {Koopman}, {Koster}, {Kuniyoshi}, {Kramer},
  {Kuper}, {Lambropoulos}, {Law}, {van Leeuwen}, {Lemaitre}, {Loose}, {Maat},
  {Macario}, {Markoff}, {Masters}, {McFadden}, {McKay-Bukowski}, {Meijering},
  {Meulman}, {Mevius}, {Middelberg}, {Millenaar}, {Miller-Jones}, {Mohan},
  {Mol}, {Morawietz}, {Morganti}, {Mulcahy}, {Mulder}, {Munk}, {Nieuwenhuis},
  {van Nieuwpoort}, {Noordam}, {Norden}, {Noutsos}, {Offringa}, {Olofsson},
  {Omar}, {Orr{\'u}}, {Overeem}, {Paas}, {Pandey-Pommier}, {Pandey}, {Pizzo},
  {Polatidis}, {Rafferty}, {Rawlings}, {Reich}, {de Reijer}, {Reitsma},
  {Renting}, {Riemers}, {Rol}, {Romein}, {Roosjen}, {Ruiter}, {Scaife}, {van
  der Schaaf}, {Scheers}, {Schellart}, {Schoenmakers}, {Schoonderbeek},
  {Serylak}, {Shulevski}, {Sluman}, {Smirnov}, {Sobey}, {Spreeuw}, {Steinmetz},
  {Sterks}, {Stiepel}, {Stuurwold}, {Tagger}, {Tang}, {Tasse}, {Thomas},
  {Thoudam}, {Toribio}, {van der Tol}, {Usov}, {van Veelen}, {van der Veen},
  {ter Veen}, {Verbiest}, {Vermeulen}, {Vermaas}, {Vocks}, {Vogt}, {de Vos},
  {van der Wal}, {van Weeren}, {Weggemans}, {Weltevrede}, {White}, {Wijnholds},
  {Wilhelmsson}, {Wucknitz}, {Yatawatta}, {Zarka}, {Zensus}, \& {van
  Zwieten}}]{lofar}
{van Haarlem}, M.~P., {Wise}, M.~W., {Gunst}, A.~W., {et~al.} 2013, \aap, 556,
  A2

\bibitem[{{Varshalovich} {et~al.}(1981){Varshalovich}, {Khersonskii}, \&
  {Syunyaev}}]{vashalovichetal81}
{Varshalovich}, D.~A., {Khersonskii}, V.~K., \& {Syunyaev}, R.~A. 1981,
  Astrophysics, 17, 273

\bibitem[{{von Hoerner}(1960)}]{vonhoerner60}
{von Hoerner}, S. 1960, \zap, 50

\bibitem[{{von Hoerner}(1963)}]{vonhoerner63}
{von Hoerner}, S. 1963, \zap, 57

\bibitem[{{Woosley} {et~al.}(2002){Woosley}, {Heger}, \& {Weaver}}]{WH2002}
{Woosley}, S.~E., {Heger}, A., \& {Weaver}, T.~A. 2002, Reviews of Modern
  Physics, 74, 1015

\bibitem[{{Woosley} \& {Weaver}(1995)}]{WW1995}
{Woosley}, S.~E. \& {Weaver}, T.~A. 1995, \apjs, 101, 181

\bibitem[{{Wouthuysen}(1952)}]{wout}
{Wouthuysen}, S.~A. 1952, \aj, 57, 31

\bibitem[{{Yoshida} {et~al.}(2003){Yoshida}, {Abel}, {Hernquist}, \&
  {Sugiyama}}]{Yoshida2003}
{Yoshida}, N., {Abel}, T., {Hernquist}, L., \& {Sugiyama}, N. 2003, \apj, 592,
  645

\end{thebibliography}
%--------------------------
%\appendix
%\section[]{}
\end{document}